\documentclass[twocolumn,groupedaddress,superscriptaddress,amsmath,amssymb,floatfix,aps,pre,showpacs]{revtex4-1}

\pdfoutput=1

\usepackage{graphicx}
\usepackage{dcolumn}
\usepackage{bm}
\usepackage[version=3]{mhchem}
\usepackage{mathtools}
\usepackage{hyperref}
\usepackage[usenames, dvipsnames]{color}
\usepackage{color}      
\usepackage{hyphenat}
\usepackage[normalem]{ulem}
\usepackage{soul}

\def\tr{{\tau_{\rm r}}}
\def\tc{{\tau_{\rm c}}}
\def\tL{{\tau_{\rm L}}}

\def\xT{X_{\rm T}}
\def\RT{R_{\rm T}}
\def\LT{{L}}
\def\mN{\overline{N}}
\def\Neff{\overline{N}_{\rm eff}}
\def\NI{\overline{N}_{\rm I}}

\def\snr{{signal-to-noise ratio}}

\def\ph{\hat{p}_{\tau_r}}
\def\phL{{\hat{p}_{{\tau_r|\LT}}}}
\def\dg{\tilde{g}_{L \to p_{\tau_{\rm r}}}}
\def\g{{g_{L \to p}}}
\def\dgs{\tilde{g}}

\def\dprime{{\prime\prime}}

\newcommand{\avg}[1]{\langle #1\rangle}
\newcommand{\bs}[1]{\boldsymbol #1}

\newcommand{\fref}[1]{Fig.~\ref{fig:#1}}

\newcommand{\flabel}[1]{\label{fig:#1}}
\newcommand{\eref}[1]{Eq.~\ref{eqn:#1}}

\newcommand{\erefstwo}[2]{Eqs.~\ref{eqn:#1}~and~\ref{eqn:#2}}
\newcommand{\erefsthree}[3]{Eqs.~\ref{eqn:#1}, ~\ref{eqn:#2}~and~\ref{eqn:#3}}
\newcommand{\erefsrange}[2]{Eqs.~\ref{eqn:#1}-\ref{eqn:#2}}
\newcommand{\elabel}[1]{\label{eqn:#1}}

\begin{document}

\preprint{AIP/123-QED}

\title[]{Theory for the optimal detection of time-varying signals in cellular sensing systems}

\author{G. Malaguti}

\author{P.R. ten Wolde}%

\affiliation{ 
AMOLF, Science Park 104, 1098 XG Amsterdam, The Netherlands
}%

\date{\today}

\begin{abstract}
  Living cells often need to measure chemical concentrations that vary
  in time. To this end, they deploy many resources such as receptors,
  downstream signaling molecules, time and energy. Here, we present a
  theory for the optimal design of a large class of sensing systems
  that need to detect time-varying signals, a receptor driving a
  push-pull network. The theory is based on the concept of the dynamic
  input-output relation, which describes the mapping between the
  current ligand concentration and the average receptor occupancy over
  the past integration time. This concept is used to develop the idea
  that the cell employs its push-pull network to estimate the receptor
  occupancy and then uses this estimate to infer the current ligand
  concentration by inverting the dynamic input-output relation.
  The theory reveals that the sensing error
  can be decomposed into two terms: the sampling error in the estimate of the
  receptor occupancy and the dynamical error that arises because the
  average ligand concentration over the past integration time may not
  reflect the current ligand concentration. The theory generalizes the
  design principle of optimal resource allocation previously
  identified for static signals, which states that in an optimally
  designed sensing system the three fundamental resource classes of
  receptors and their integration time, readout molecules, and energy are equally limiting so that no resource is
  wasted. However, in contrast to static signals, receptors and power
  cannot be traded freely against time to reach a desired sensing
  precision: there exists an optimal integration time that maximizes
  the sensing precision, which depends on the number of receptors, the
  receptor correlation time, and the correlation time and variance of
  the input signal. Applying our theory to the chemotaxis system of
  {\it Escherichia coli} indicates that this bacterium has evolved to
  optimally sense shallow gradients.
\end{abstract}

\maketitle

\section{\label{sec:Intro}Introduction}

Living cells often need to sense and respond to chemical
signals that vary in time. This is particularly true for cells
that navigate through their environment. Interestingly, experiments
have revealed that cells can measure chemical concentrations with high
precision \cite{Berg:1977bp,Sourjik:2002fk,Ueda2007}. This raises the
question how accurately cells can measure time-varying signals.

Cells measure chemical concentrations via receptors on their
surface. These measurements are inevitably corrupted by the stochastic
arrival of the ligand molecules by diffusion and by the stochastic
binding of the ligand to the receptor. Berg and Purcell pointed out
that cells can increase the number of measurements to reduce the
sensing error in two principal ways \cite{Berg:1977bp}. One is to
simply increase the number of receptors. The other is to take more
measurements per receptor; here, the cell infers the concentration not
from the instantaneous number of ligand-bound receptors, but rather
from the average receptor occupancy over an integration time
\cite{Berg:1977bp}.  While many studies have addressed the question
how time integration sets the fundamental limit to the precision of
sensing static concentrations
\cite{Bialek2005,Wang2007,Rappel2008,Endres2009,Hu2010,Mora2010,Mehta2012,Govern2012,Govern2014,Govern2014b,Kaizu2014,Mugler:2016dy,Fancher:2017ba}
(for review, see \cite{tenWolde:2016ih}), how accurately time
integration can be performed for time-varying signals is a wide-open
question \cite{Becker2015}. A theory that can describe how the
  sensing precision depends on the design of the system and
  predict what the optimal design is that maximizes the sensing
  precision is lacking.

  Biochemical networks that implement the mechanism of time
  integration require cellular resources to be built and run.
  Receptors and time are needed to take the concentration
  measurements, downstream molecules are necessary to store the
  ligand-binding states of the receptor in the past, and energy is
  required to store these states reliably. In a previous study on
  sensing static signals that do not vary on the timescale of the
  cellular response, we showed that three resource classes---receptors
  and their integration time, readout molecules, and
  energy---fundamentally limit sensing like weak links in a chain
  \cite{Govern2014}. This yields the design principle of optimal
  resource allocation, which states that in an optimally designed
  system each resource class is equally limiting so that no resource
  is in excess \cite{Govern2014}.  Within these
  classes, resources can be traded freely against each other: time can
  not only be traded against receptors---a system consisting of one
  receptor that takes many measurements over time can reach the same
  sensing precision as one containing many receptors that take one
  measurement each---but also against power---many noisy measurements
  can provide the same information as one reliable measurement.

Cells live, however, in a highly dynamic environment and they
  often respond on a timescale that is comparable to that on which the
  input signal varies. Examples are cells (or nuclei) that during
  embryonic development differentiate in response to time-varying
  morphogen gradients \cite{Durrieu:2018kj} or cells that navigate
through their environment
\cite{Tostevin2009,Sartori:2011fh,Long:2016ub};  these cells shape, via their movement, the
statistics of the input signal,
creating a correlation time of the input signal that is comparable to
the timescale of the response. In these scenarios, the accuracy of
sensing depends not only on properties of the cellular sensing system, but
also on the dynamics of the input
signal.  It is indeed far from clear
whether the design principles uncovered for systems sensing static
concentrations \cite{Govern2014} also hold for those that need to
detect time-varying signals. In particular, for sensing time-varying
signals we expect that time itself becomes a fundamental resource. A
longer integration time will not only reduce the receptor noise but
also distort the input signal \cite{Becker2015,Monti:2018ee}. This
raises many questions: Can the design principle of optimal resource
allocation be generalized to time-varying signals? If so, what does it
predict for the optimal design of the system? How does that depend on
the statistics of the input signal? In particular, how does the power
and the number of receptor and readout molecules required to maintain
a desired sensing precision depend on the timescale and the strength
of the input fluctuations?

To address these questions we present a theory for the optimal design
of cellular sensing systems that need to measure time-varying ligand
concentrations.  The theory applies to the large class of systems in
which a receptor drives a push-pull network \cite{Goldbeter1981}.
These systems are omnipresent in prokaryotic and eukaryotic cells
\cite{Alon:2007tz}. Examples are GTPase cycles, as in the Ras system,
phosphorylation cycles, as in MAPK cascades, and two-component systems
like the chemotaxis system of {\em Escherichia coli}. These systems
employ the mechanism of time integration, in which the ligand
concentration is inferred from the average receptor occupancy over the
past integration time \cite{Govern2014}. We thus do not consider the
sensing strategy of maximum-likelihood estimation, in which the
concentration is estimated from the duration of the unbound receptor
state
\cite{Endres2009,Mora2010,Lang:2014ir,Hartich:2016gq,tenWolde:2016ih}.

To develop a unified theory of sensing, we combine ideas on
  information transmission via time-varying signals from
  Refs. \cite{Tostevin2010,Hilfinger:2011ev,Bowsher2013} with the
  sampling framework from Ref. \cite{Govern2014}. Our theory is based
  on a new concept, the {\em dynamic} input-output relation
  $p_\tr(\LT)$, which describes the mapping between the average
  receptor occupancy $p_\tr$ over the past integration time $\tr$ and
  the current concentration $\LT$; it differs fundamentally from the
  conventional static input-output relation, because it takes into
  account the dynamics of the input signal and the finite response
  time of the system. The dynamic input-output relation allows us to
  develop the notion that the cell employs its push-pull network to
  estimate the receptor occupancy and then uses this estimate to infer
  the current concentration, by inverting $p_\tr (\LT)$.  Our theory
reveals that the sensing error can be decomposed into two terms, which
each have a clear intuitive interpretation. One term, the sampling
error, describes the sensing error that arises from the finite
accuracy by which the receptor occupancy is estimated. This error
depends on the number of receptor samples as set by the number of
readout molecules and the integration time; their independence as
given by the receptor-sampling interval and the receptor-ligand
correlation time; and their reliability as determined by fuel
turnover. The other term is the dynamical error, and is related to the
error introduced in \cite{Bowsher2013}. This error is determined by
how much the concentration in the past integration time reflects the
current concentration that the cell aims to estimate; it depends
besides the integration time on the timescale on which the input
varies.

Our theory gives a comprehensive view on the optimal design of a
cellular sensing system. Firstly, it reveals that the resource
allocation principle of \cite{Govern2014} can be generalized to
time-varying signals. There exist three fundamental resource
classes---receptors and their integration time, readout molecules, and
power and integration time---which each fundamentally limit the
accuracy of sensing; and, in an optimally designed system, each
resource class is equally limiting the sensing precision. The optimal
resource allocation principle thus gives the relationship between
receptors, integration time, readout molecules, and power so that none
of these cellular resources is in excess and thus wasted.  However, in
contrast to sensing static signals, time cannot be freely traded
against the number of receptors and the power to achieve a desired
sensing precision: there exists an optimal integration time that
maximizes the sensing precision, which arises as a trade-off between
the sampling error and the dynamical error. This optimal integration
time depends on the statistics of the input signal and on the number
of receptors. Together with the resource allocation principle it
completely specifies the optimal design of the system in terms of its
resources protein copies, time, and energy.

Our theory also makes a number of specific predictions. The optimal
integration time decreases as the number of receptors is increased,
because this allows for more instantaneous measurements. It also
decreases when the input signal varies more rapidly and/or more
strongly. Moreover, our allocation principle reveals that when the
input signal varies more rapidly, both the number of receptors and the
power must increase to maintain a desired sensing precision, while
the number of readout molecules does not. Finally, we test our
prediction for the optimal integration time for the chemotaxis system
of {\it Escherichia coli}; this analysis indicates that the chemotaxis
system has evolved to optimally sense shallow concentration gradients.

\section{\label{sec:model}Theory}

\subsection{The set up of the problem}
We consider a single cell that needs to sense a time-varying ligand
concentration $\LT(t)$, see \fref{model}(a). The ligand concentration
dynamics is modeled as a stationary Markovian signal specified by the
mean (total) ligand concentration $\overline{L}$, the variance
$\sigma^2_L$ and the correlation time $\tL=\lambda^{-1}$, which sets
the timescale of the input fluctuations. It obeys Gaussian statistics
\cite{Tostevin2010}. 

The concentration is measured via $\RT$ receptor proteins on the cell
surface, which independently bind the ligand~\cite{tenWolde:2016ih},
$\ce{L + R <=>[k_1][k_2] RL}$. The correlation time of the receptor
state is given by $\tc=1/(k_1 \overline{L}+k_2)$. It
determines the timescale on which independent concentration
measurements can be made. Denoting the average number of ligand-bound
receptors as $\overline{RL}$, the receptor occupancy is
$p=\overline{RL}/\RT=k_1 \overline{L}\tc$. This shows that for
a given $p$ the correlation time $\tc = p/(k_1\overline{L})=\mu^{-1}$
is fundamentally bounded by the ligand concentration
$\overline{L}$ and the ligand diffusion constant, which limits
the binding rate $k_1$ \cite{Berg:1977bp,Bialek2005,Kaizu2014,tenWolde:2016ih}.

The ligand-binding state of the receptor is read out via a push-pull
network \cite{Goldbeter1981}, which is a common non-equilibrium
signaling motif in prokaryotic and eukaryotic cells
\cite{Alon:2007tz}. In this system, fuel turnover is used to drive the
chemical modification of a downstream readout protein $x$, see
\fref{model}(b). The most common scheme is phosphorylation fueled by
the hydrolysis of adenosine triphosphate (ATP). The receptor, or an
enzyme associated with it such as CheA in {\it E. coli}, catalyzes the
modification of the readout, $\ce{x + RL + ATP <=> x^* + RL +
  ADP}$. The active readout proteins $x^*$ can decay spontaneously or
be deactivated by an enzyme, such as CheZ in {\it E. coli}, $\ce{x^*
  <=>x + Pi}$. Inside the living cell the system is maintained in a
non-equilibrium steady state by keeping the concentrations of ATP, ADP
(adenosine diphosphate) and Pi (inorganic phosphate) constant. We
absorb their concentrations and the activities of the kinase and, if
applicable, phosphatase in the (de)phosphorylation rates,
coarse-graining the modification into instantaneous second order
reactions: $\ce{x + RL <=>[k_{\rm f}][k_{-{\rm f}}] x^* + RL}$,
$\ce{x^* <=>[k_{\rm r}][k_{-{\rm r}}]x + Pi}$. This system has a
relaxation time $\tr=1/[(k_{\rm f}+k_{-{\rm f}})\overline{RL}+k_{\rm
  r}+k_{-{\rm r}}]=\mu^{\prime -1}$. It sets the lifetime of the active readout
molecules, which determines how long these molecules can carry
information on the ligand binding state of the receptor. The
relaxation time $\tr$ thus sets the integration time of the receptor
state.

The  deviations of $RL$ and $x^*$ away from their steady-state values are
given by (see section S-I of the {\it SI}):
\begin{align}
  \delta \dot{RL}(t)  &= \rho \delta \LT(t)-\mu \delta RL(t)+\eta_{RL} \elabel{dRL_main},\\
  \delta \dot{x^*}(t) &= \rho^\prime \delta RL(t)-\mu^\prime \delta x^*(t) + \eta_{x^*},\elabel{ddx_main}
\end{align}
where $\rho$ and $\rho^\prime$ are functions of the rate
constants and the noise terms $\eta_{RL},\eta_{x^*}$ model the
noise in receptor-ligand binding and readout phosphorylation,
respectively.

\begin{figure}
  \centering
  \includegraphics[width=0.48\textwidth]{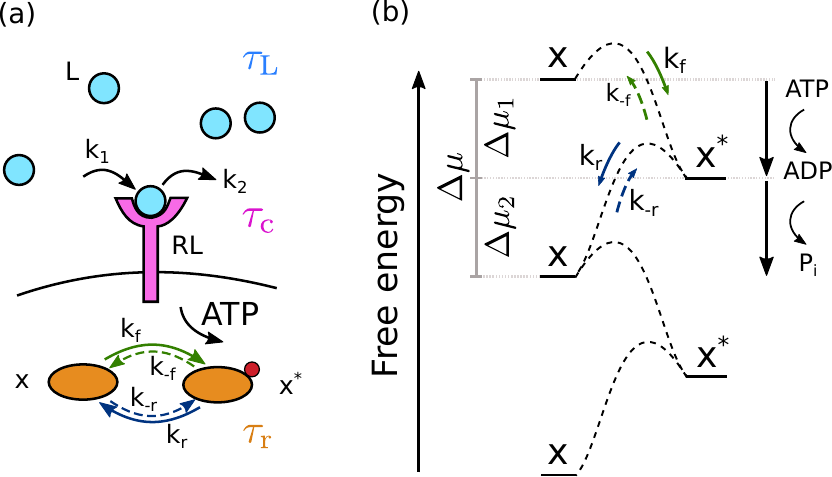}
  \caption{\label{fig:model}\textbf{The cell signaling network.} (a)
    The time-varying ligand concentration is modeled as a
    Markovian signal with mean $\overline{L}$, variance
      $\sigma^2_L$ and correlation time $\tL=\lambda^{-1}$.
    A free ligand molecule L (light blue circle) can
    bind at rate $k_1$ to a free receptor R (magenta protein) on the
    cell membrane (black line), forming the complex RL, and unbind at
    rate $k_2$ from RL. The correlation time of the receptor state
    is $\tc$. The complex RL catalyzes the phosphorylation
    reaction, driven by ATP conversion, of a downstream readout from
    the unphosphorylated (inactive) state $x$ to the phosphorylated
    (active) state $x^*$, with rate $k_{\rm f}$. The phosphorylated
    readout then spontaneously decays to the $x$ state with rate
    $k_{\rm r}$. Microscopic reverse reactions of each signaling
    pathway are represented by dashed arrows. The relaxation time
    of the push-pull network is $\tr$. (b) Free-energy landscape of a
    readout molecule across the activation/deactivation reactions.
    Fuel turnover, provided by ATP conversion, drives the
    activation (phosphorylation) reaction characterized by the forward rate
    $k_{\rm f}$ and its microscopic reverse rate $k_{\rm -f}$ (green
    arrows). Associated with this activation reaction is a free-energy
    drop $\Delta \mu_1=\log\frac{k_{\rm f}\overline{x}}{k_{-{\rm
          f}}\overline{x}^*}$. The deactivation pathway corresponds to
    the spontaneous release of the inorganic phosphate; it is
    characterized by the rate $k_{\rm r}$ and its microscopic reverse
    $k_{\rm -r}$ (blue arrows) and corresponds to a free-energy drop
    $\Delta \mu_2=\log\frac{k_{\rm r}\overline{x}^*}{k_{-{\rm
          r}}\overline{x}}$.}
\end{figure}

\subsection{\label{sec:SNRdef} The cell sensing precision}

\noindent{\bf Signal-to-noise ratio}\\
The time-varying ligand concentration has a distribution of
instantaneous values $\LT(t)$, and we would like to know how many of
these the system can resolve. To this end we define the
signal-to-noise ratio (SNR), see \fref{theory}(a):
\begin{align}
\text{SNR} \vcentcolon = \frac{\sigma^2_{\LT}}{(\delta \hat{\LT})^2}.
\elabel{SNRdef}
\end{align}
Here $\sigma^2_{\LT}$ is the variance of the ligand concentration
$\LT(t)$; it is a measure for the total number of input states. 
The quantity $(\delta \hat{\LT})^2$ is the error in the estimate of the
current ligand concentration.
The \snr\ thus quantifies the number of distinct ligand concentrations
that the system can resolve. Since the system is stationary and time invariant, we can omit the argument in $\LT(t)$ and write $\LT=\LT(t)$.\\

\begin{figure*}
  \centering
  \includegraphics[width=0.97\textwidth]{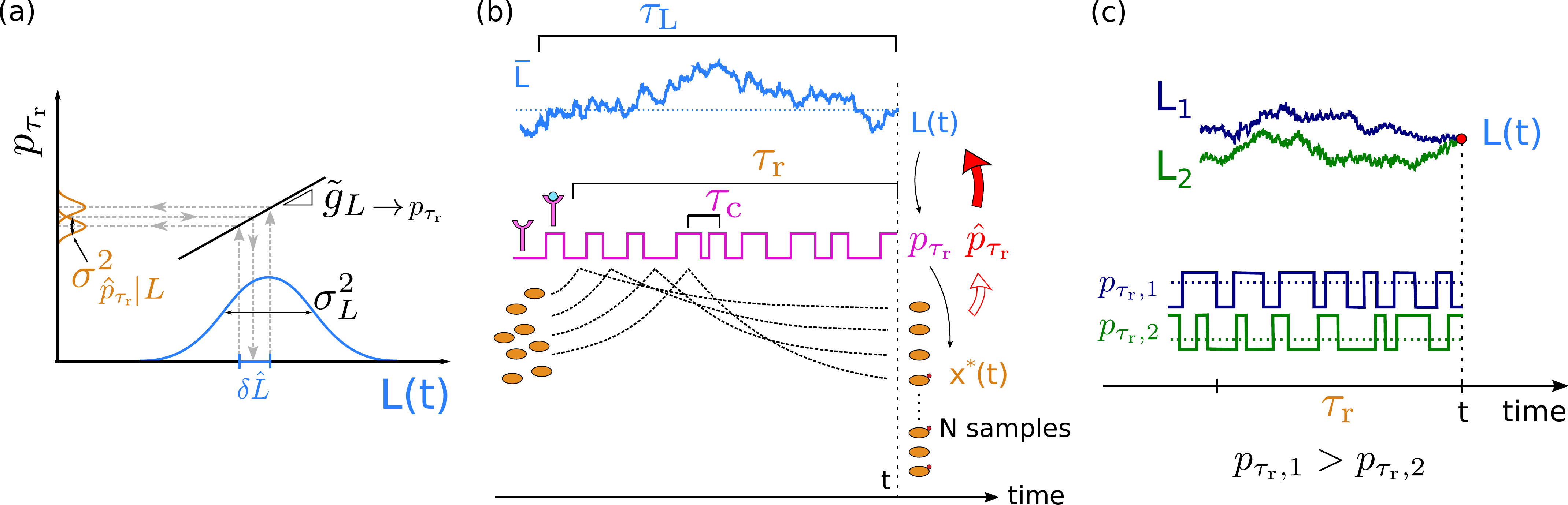}
  \caption{\label{fig:theory}\textbf{The precision of estimating a
      time-varying ligand concentration $\LT(t)$.}  (a) The cell
    estimates the current ligand concentration $\LT(t)$ by estimating the
    average receptor occupancy $p_\tr$ over the past
    integration time $\tr$ and by inverting the dynamic input-output
    relation $p_\tr(\LT)$. The error in the estimate of the
    concentration $(\delta \hat{L})^2=\sigma^2_{{\ph}|\LT} / \dg^2$
    depends on the variance $\sigma^2_{\ph|\LT}$ in the estimate of
    the average receptor occupancy $\ph$ and the dynamic gain $\dg$,
    the slope of $p_\tr(\LT)$, which determines how the error in $\ph$
    propagates to that in $\delta \hat{L}$. The input distribution has
    width $\sigma^2_\LT$. (b) The average receptor occupancy $p_\tr$
    over the past integration time $\tr$ is estimated via the
    downstream network, which is modelled as a device that discretely
    samples the ligand-binding state of the receptor via its readout
    molecules $x$ \cite{Govern2014}; the fraction of modified readout
    molecules provides an estimate of $p_\tr$, see \eref{ptr}. The
    sensing error has two contributions, \eref{minfo_final_sampl}: the
    sampling error and the dynamical error. The sampling error arises
    from the error in the estimate of $p_\tr$ that is due to the
    stochasticity of the sampling process; it depends on the number of
    samples, their independence and their accuracy.  (c) The dynamical
    error arises because the current ligand concentration $\LT(t)$ is
    estimated via the average receptor occupancy $p_\tr$ over the past
    integration time $\tr$: the latter depends on the ligand
    concentration in the past $\tr$, which will, in general, deviate
    from the current concentration.  Two different input trajectories
    ($L_{1}$ in blue, $L_{2}$ in green) ending at time $t$ at the
    \textit{same} value $\LT(t)$ (red dot) lead to different estimates
    of $\LT(t)$, due to their different average receptor occupancy
    ($p_{\tr,1}>p_{\tr,2}$) in the past $\tr$.  }
\end{figure*}

\noindent {\bf Inferring concentration from readout}\\ The cell
estimates the current ligand concentration $\LT(t)$ from the
instantaneous number of active readout molecules $x^*(t)$. In the
Gaussian model employed here, we can calculate the SNR defined by
\eref{SNRdef}, and it is related to the mutual information $I(x^*;\LT)
= 1/2 \ln (1+{\rm SNR})$ between the input $\LT$ and the output
$x^*$ (section S-II {\it SI}). However, the resulting expression for
the SNR is not very instructive (Eq. S21):
\begin{align}
{\rm SNR}^{-1}  &= \frac{(\lambda+\mu)^2(\lambda+\mu')^2}{\rho^2\rho^{\prime^2}\sigma^2_\LT} f(1-f)X_{\rm T} \nonumber \\
  &+ \frac{(\lambda+\mu)^2(\lambda+\mu')^2}{\sigma^2_\LT \mu^\prime(\mu+\mu^\prime)\rho^2} p(1-p)\RT \nonumber \\
  & +\frac{(\lambda+\mu)(\lambda+\mu')(\lambda+\mu+\mu^\prime)}{\mu
    \mu^\prime(\mu+\mu^\prime)} -1.\elabel{LNA}
\end{align} 

This approach cannot elucidate the design logic of the system, because
it treats the signal transmission from the input $\LT$ to the output
$x^*$ as a black box. The central quantity in this calculation is the
covariance $\sigma^2_{\LT,x^*}$ between the ligand $\LT(t)$ and the
readout $x^*(t)$ (see section S-II), which does not reveal how the
signal is relayed from the input to the output. To elucidate the
system's design principles, we have to open the black box: we need to
recognize that the input signal is transmitted to the output via the
receptor, and that the cell does not estimate the ligand concentration
from $x^*$ directly, but rather via its receptor (see
\fref{model}). We can indeed arrive at a much more illuminating form
of the same result as \eref{LNA} by starting from the notion that
the cell uses its readout system to estimate the receptor occupancy,
from which the ligand concentration is then inferred. However, to
develop this notion into a theory, we need new
concepts, which we describe next.\\

\noindent {\bf Inferring concentration from
  receptor occupancy} \\
The central idea of our theory is illustrated in \fref{theory}a: the
cell employs the push-pull network to estimate the average receptor
occupancy $p_\tr$ over the past integration time $\tr$, and then uses
this estimate $\phL$ to infer the current concentration $\LT$ by
inverting the mapping $p_\tr(\LT)$. The sensing error is then
determined by how accurately the cell estimates $p_\tr$ and how the
error in $\phL$ propagates to that in $\LT$, which is determined by
$p_\tr(\LT)$. We now first give an overview of the central concepts.

{\bf Dynamic input-output relation} The mapping $p_\tr(\LT)$ is the
{\em dynamic input-output relation}. It gives the average receptor
occupancy over the past integration time $\tr$ {\em given} that the
{\em current} value of the input signal is $\LT=\LT(t)$, see
\fref{theory}(a). Here, the average is not only over the noise in
receptor-ligand binding and readout activation (\fref{theory}(b)), but
also over the subensemble of past input trajectories that each end at
the same current concentration $\LT$ (\fref{theory}(c))
\cite{Tostevin2010,Hilfinger:2011ev,Bowsher2013}. In contrast to the
conventional, static input-output relation $p(\LT_{\rm s})$, which
gives the average receptor occupancy $p$ for a steady-state ligand
concentration $\LT_{\rm s}$ that does not vary in time, the dynamic
input-output relation takes into account the dynamics of the input
signal and the finite response time of the system. It depends on all
the timescales in the problem: the timescale of the input, $\tL$, the
receptor-ligand correlation time $\tc$, and the integration time
$\tr$. Only when $\tL\gg \tc,\tr$, does
the dynamic input-output relation $p_\tr(\LT)$ become equal to the
static input-output relation $p(\LT_{\rm s})$.

{\bf Sensing error} Linearizing the dynamic input-output relation
$p_\tr(\LT)$ around the mean ligand concentration $\overline{\LT}$ (see
\fref{theory}a) and
using the rules of error propagation, the expected error in the
concentration estimate is then
\begin{align}
(\delta \hat{\LT})^2 = \frac{\sigma^2_{\phL}}{\dg^2}.
\elabel{LTerr}
\end{align}
Here $\sigma^2_{\ph|\LT}$ is the variance in the estimate $\phL$ of
the average receptor occupancy over the past $\tr$ {\em given} that
the current input signal is $\LT$, see \fref{theory}(a). The quantity
$\dg$ is the {\em dynamic} gain, which is the slope of the dynamic
input-output relation $p_\tr(\LT)$; it determines how much an error in
the estimate of $p_\tr$ propagates to that in $\LT$. \eref{LTerr}
generalizes the expression for the error in sensing static
concentrations
\cite{Berg:1977bp,Bialek2005,Wang2007,Mehta2012,Kaizu2014,Govern2014,tenWolde:2016ih}
to that of time-varying concentrations.

{\bf SNR} Combining \erefstwo{LTerr}{SNRdef} yields the signal-to-noise ratio:
\begin{align}
\text{SNR} = 
\frac{\tilde{g}^2_{\LT \to p_{\tr}}}{\sigma^2_{\phL}}\; \sigma^2_{\LT}.
\elabel{SNRp}
\end{align}

{\bf Estimating the receptor occupancy} To derive the error in
estimating $p_\tr$, $\sigma^2_\phL$, we view, following our earlier
work \cite{Govern2014}, the push-pull network as a device that
discretely samples the receptor state (see \fref{theory}(b)). The
principle is that cells employ the activation reaction ${x + RL \to
  x^* + RL}$ to store the state of the receptor in stable chemical
modification states of the readout molecules. Readout molecules that
collide with a ligand-bound receptor are modified, while those that
collide with an unbound receptor are not (\fref{theory}(b)).
The readout molecules serve as samples of the receptor at the time
they were created, and collectively they encode the history of the
receptor.  The average receptor occupancy $p_\tr$ over the past
integration time $\tr$ is thus estimated from the current number of
active readout molecules $x^*(\LT(t))=x^*(L)$:
\begin{align}
\phL = \frac{x^*(L) }{\mN},
\elabel{ptr}
\end{align}
where $\mN$ is the average number of samples obtained during $\tr$.
To determine the effective number of independent samples, we need to
consider not only the creation of the samples, but also their decay
and accuracy. Samples decay via the deactivation reaction ${x^* \to
  x}$, which means that they only provide information on the receptor
occupancy over the past $\tr$.  In addition, both the activation and
the deactivation reaction can happen in their microscopic reverse
direction, which corrupts the coding.  Energy is needed to break time
reversibility and protect the coding. Furthermore, for time-varying
signals, we also need to recognize that the samples correspond to the
ligand concentration over the past integration time $\tr$, which will
in general differ from the current concentration $\LT$ that the
cell aims to estimate.  While a finite $\tr$ is necessary for time
integration, it will also lead to a systematic error in the estimate
of the concentration that the cell cannot reduce by taking more
receptor samples.

{\bf Estimating concentration from ${\bs p_\tr}$ is no
  different from that via readout ${\bs x^*}$} Because the average number of
samples $\mN$ is
a constant, it follows from \eref{ptr} that the variance in $x^*$
given an input $\LT$ is $\sigma^2_{x^*|\LT} = \sigma^2_{\phL} \mN^2$
while the gain from $\LT$ to $x^*$ is $\tilde{g}^2_{\LT \to x^*} =
\tilde{g}^2_{\LT \to p_\tr} \mN^2$. Consequently, the absolute error
$(\delta \hat{L})^2$ in estimating the concentration via $x^*$, $(\delta
\hat{L})^2 = \sigma^2_{x^*|\LT} / \tilde{g}^2_{\LT \to x^*}$, is the same as
that of \eref{LTerr}\,: because the instantaneous number of active
readout molecules $x^*$ reflects the average receptor occupancy
$p_\tr$ over the past $\tr$, estimating the ligand concentration from
$x^*$ is no different from inferring it from the average receptor
occupancy $\phL = x^* / \mN$. In the {\it Supporting Information} we
show explicitly that the central result of our manuscript that follows
from the sampling framework, \eref{minfo_final_sampl}, is indeed
identical to that of \eref{LNA} (section S-IV).

{\bf Key steps derivation central result} We now sketch the derivation
of the central result for a simpler system, the irreversible network
($k_{\rm -f} =k_{\rm -r}=0$). For details and the result on the full
system, see {\it SI}.

{\bf Dynamic gain} The dynamic gain $\dg=\delta p_\tr / \delta \LT(t)$
quantifies the mapping between the deviation $\delta \LT(t)\equiv
\LT(t) - \overline{\LT}$ of the current ligand concentration $\LT(t)$
from its mean $\overline{\LT}$ and the deviation $\delta p_\tr$ of the
average receptor occupancy over the past integration time $\tr$ from
its mean $p$, see
\fref{theory}(a). 
This average is taken by the readout molecules at time $t$. Taking
into account deactivation, the probability that a readout molecule at
time $t$ provides a sample of the receptor at an earlier time $t_i$ is
$p(t_i|{\rm sample})= e^{-(t-t_i)/\tr}/\tr$
\cite{Govern2014}. Averaging the receptor occupancy over the sampling
times $t_i$ then yields
\begin{align}
\delta p_\tr=\int_{-\infty}^tdt_i \avg{\delta
  n(t_i)}_{\delta \LT(t)} \frac{e^{-(t-t_i)/\tr}}{\tr}.
\elabel{ntr_main}
\end{align}
Here, $\avg{\delta n(t_i)}_{\delta \LT(t)}= \avg{n(t_i)}_{\delta
  \LT(t)} - p$ is the average deviation in the receptor occupancy
$n(t_i)=0,1$ at time $t_i$ {\em given that the ligand concentration at time
$t$ is $\delta \LT(t)$}, where the average is taken over
receptor-ligand binding noise and the subensemble of ligand trajectories
that each end at $\delta \LT(t)$ (see \fref{theory}c). We can compute
it within the linear-noise approximation:
\begin{align}
\avg{\delta n(t_i)}_{\delta \LT(t)}= \rho_n \int_{-\infty}^{t_i} dt^\prime
\avg{\delta \LT(t^\prime)}_{\delta \LT(t)} e^{-(t_i -
    t^\prime)/\tc},
\elabel{nti_main}
\end{align}
where $\rho_n = p(1-p) / (\overline{L}_T \tc)$ and $\avg{\delta
  \LT(t^\prime)}_{\delta \LT(t)}$ is the average ligand concentration at
time $t^\prime$ given that the concentration at time $t$ is
$\delta \LT(t)$. It is given by
\cite{Bowsher2013}
\begin{align}
\avg{\delta \LT(t^\prime)}_{\delta \LT(t)}= \delta \LT(t)
e^{-|t-t^\prime|/\tL}.
\elabel{Ltp_main}
\end{align}
Combining \erefsrange{ntr_main}{Ltp_main} yields 
\begin{align}
\dg&= \frac{p(1-p)}{\overline{L}}\left(1+\frac{\tc}{\tL}\right)^{-1}\left(1+\frac{\tr}{\tL}\right)^{-1},
\elabel{dg_main}\\
&=g_{\LT \to p}\left(1+\frac{\tc}{\tL}\right)^{-1}\left(1+\frac{\tr}{\tL}\right)^{-1}.
\end{align}
The dynamic gain $\dg$ depends on all the timescales in the problem. Only
when $\tL \gg \tr, \tc$ is the average ligand concentration over the
subensemble of trajectories ending at $\delta \LT(t)$ equal to current
concentration $\delta \LT(t)$ (see \fref{theory}(c)), and does $\dg$
become equal to its maximal value, the static gain $g_{\LT \to
  p}=p(1-p) / \overline{L}$.

{\bf The error in estimating the receptor occupancy} Using the law of
total variance, the error $\sigma^2_{\ph|\LT}$ in the estimate of the
receptor occupancy $p_\tr$ over the past integration time $\tr$ is
given by
\begin{align}
\elabel{noiseaddv1_main}
\sigma_{\ph|\LT}^2 =    \text{var} \left[ E  (\phL |N) \right] + E
\left[ \text{var}(\phL | N) \right].
\end{align}
The first term reflects the variance of the mean of $\phL$ given the
number of samples $N$; the second term reflects the mean of the
variance in $\phL$ given the number of samples $N$
\cite{Govern2014}. 

The first term of \eref{noiseaddv1_main} is given by (see Eq. S48)
\begin{align}
 \text{var} \left[ E  (\phL |N) \right] = \frac{p^2}{\mN}\elabel{varMeanN_main}.
\end{align}
This contribution reflects the fact that with a push-pull
network as considered here, the cell cannot discriminate between those
readout molecules that have collided with an unbound receptor, and
hence provide a sample of the receptor, and those that have not
collided with a receptor at all; this term is zero for a
bifunctional kinase where the unbound receptor catalyzes readout
deactivation  \cite{Govern2014}.

The second term of \eref{noiseaddv1_main} contains two
contributions. First we note that
(Eq. S50 and Appendix S-B)
\begin{align}
&{\rm var}\left(\frac{\sum_{i=1}^N n_i(t_i)}{N}|N\right)\nonumber\\
&=\frac{p(1-p)}{N} + \overline{E\avg{\delta n_i(t_i)
      \delta n_j(t_j)}}_{\delta \LT(t)} - \dgs^2 \sigma^2_\LT \elabel{vardelta_main},
\end{align}
where $\delta n_i(t_i) = n_i(t_i) - p$, $E$ denotes an average over the
sampling times $t_i$, and the overline an average over $\delta \LT$. The receptor covariance $\overline{E\avg{\delta
    n_i(t_i) \delta n_j(t_j)}}_{\delta \LT(t)}$ can be decomposed into two
contributions. The first combines with the first term of
\eref{vardelta_main} to yield (Eq. S61)
\begin{align}
 E\left[ \text{var}(\phL | N) \right]^{\rm samp}= \frac{p(1-p)}{\mN_{\rm I}},\elabel{EvarNfix_main}
\end{align}
where $\mN_{\rm I}=f_I \mN$. Here, $f_I=1/(1+2\tc / \Delta)$, with
$\tc$ the receptor-ligand correlation time and $\Delta$ the spacing
between the receptor samples, is the fraction of the $\mN$ samples
that are independent.  \eref{EvarNfix_main} is the error in the
estimate of the receptor occupancy based on a single measurement---the
variance of the receptor occupancy $p(1-p)$---divided by the total
number of independent measurements, $\mN_{\rm I}=f_I \mN$.  Together
with the first term of \eref{noiseaddv1_main} (i.e. \eref{varMeanN_main})
\eref{EvarNfix_main} yields the sampling error in the estimate of the average
receptor occupancy over $\tr$:
\begin{align}
\sigma^{2,\,\, {\rm samp}}_{\phL} = \frac{p^2}{\mN} +
\frac{p(1-p)}{\mN_{\rm I}}.\elabel{SampErrp_main}
\end{align}
Both contributions to $\sigma^{2,\,\, {\rm samp}}_{\phL}$ are governed
by the nature of the receptor sampling process and do not depend on the
input statistics. They are indeed the same as those for sensing static
concentrations, derived previously \cite{Govern2014}.

The second contribution to \eref{noiseaddv1_main}  comes
from the second contribution to $\overline{E\avg{\delta
    n_i(t_i) \delta n_j(t_j)}}_{\delta \LT(t)}$ in
\eref{vardelta_main}. It combines with the third term of \eref{vardelta_main} to
yield (see Eq. S72)
\begin{align}
\sigma^{2, \,\, {\rm dyn}}_{\phL}&= E\left[ \text{var}(\phL | N) \right]^{\rm dyn}\nonumber\\
&=\dgs^2 \sigma^2_\LT \left[
  \left(1+\frac{\tc}{\tL}\right)\left(1+\frac{\tr}{\tL}\right)\left(1+\frac{\tc\tr}{\tL(\tc+\tr)}\right)-1\right]
\elabel{dynErrp_main}
\end{align}
This is the dynamical error in estimating $p_\tr$. It corresponds to
the variation in $p_\tr$ that arises from the different concentration
trajectories in the past $\tr$ that each end at $\delta \LT(t)$, see
Fig.~\ref{fig:theory}(c). This error does depend on the statistics of
the input signal: it increases with the width of the input distribution,
$\sigma^2_L$, and decreases with the input timescale $\tL$.

The error in estimating the average receptor occupancy $p_\tr$ is then
given by
\begin{align}
\sigma^2_{\phL}=\sigma^{2,\,\, {\rm samp}}_{\phL} + \sigma^{2, \,\,
  {\rm dyn}}_{\phL},\elabel{Errp_main}
\end{align}
where $\sigma^{2, \,\, {\rm samp}}_{\phL}$ and $\sigma^{2, \,\, {\rm
    dyn}}_{\phL}$ are given by \erefstwo{SampErrp_main}{dynErrp_main}.

{\bf Central result} To know how the error $\sigma^{2}_{\phL}$ in the
estimate of the average receptor occupancy $p_\tr$ propagates to the
error $(\delta \hat{L})^2$ in the estimate of the ligand concentration, we
divide \eref{Errp_main} by the dynamic gain $\dg$ given by \eref{dg_main} (see
\eref{SNRp}). For the full system, the reversible push-pull network,
this yields the central result of our manuscript, the \snr\ in terms of
the total number of receptor samples, their independence, their
accuracy, and the timescale on which they are generated:
\begin{widetext}
\begin{eqnarray}
\elabel{minfo_final_sampl}
\text{SNR}^{-1} &=&
\underbrace{\left(1+\frac{\tc}{\tL}\right)^2
  \left(1+\frac{\tr}{\tL}\right)^2\left[\frac{\left(\overline{\LT}/\sigma_\LT\right)^2}{p(1-p)\mN_{\rm
        I}} +
    \frac{\left(\overline{\LT}/\sigma_\LT\right)^2}{(1-p)^2\Neff}\right]}_\text{sampling
  error} + \underbrace{\left(1+\frac{\tc}{\tL}\right) \left(1+\frac{\tr}{\tL}\right) \left(1+\frac{\tc \tr}{\tL(\tc+\tr)}\right)-1}_\text{dynamical error}.
\end{eqnarray}
\end{widetext}
This expression represents exactly the same result as that obtained by
the straightforward linear-noise calculation, \eref{LNA}
(section S-IV). However, it is much more
illuminating.  It shows that the sensing error ${\rm SNR}^{-1}$ can be
decomposed into two distinct contributions, which each have a clear
interpretation: the {\em sampling error}, arising from the
stochasticity in the sampling of the receptor state, and the {\em
  dynamical error}, arising from the dynamics
of the input signal.

When the timescale of the ligand fluctuations $\tL$ is much longer
than the receptor correlation time $\tc$ and the integration time
$\tr$, $\tL\gg \tr,\tc$, the dynamical error reduces to zero and only
the sampling error remains. In this limit, the prefactor
$(1+\tc/\tL)^{2}(1+\tr/\tL)^{2}$ becomes unity, and the relative
sensing error $(\delta \hat{\LT}/\overline{\LT})^2$ (instead of ${\rm
  SNR}^{-1}=(\delta \hat{\LT} / \sigma_\LT)^2$) reduces to that of
estimating {\em static} concentrations, derived previously in
Ref.~\cite{Govern2014}. Here, $\Neff$ is the total number of effective
samples and $\NI$ is the number of these that are independent
\cite{Govern2014}. For the full system they are given by:
\begin{align}
\NI = \underbrace{\frac {1}{(1+2\tau_c/\Delta)}}_{f_I}
\underbrace{\overbrace{\frac{\left( e^{\beta \Delta \mu_1} -
        1 \right) \left( e^{\beta \Delta \mu_2} - 1 \right)}{ e^{\beta
        \Delta \mu}
      - 1}}^{q}\overbrace{\frac{\dot{n}
      \tau_r}{p}}^{\bar{N}} }_{\bar{N}_{\rm eff}}.
\elabel{NI}
\end{align}
The quantity $\dot{n}=k_{\rm f}p \RT \overline{x}-k_{-{\rm f}} p\RT
\overline{x}^*$ is the net flux of $x$ around the cycle of activation
and deactivation. It equals the rate at which $x$ is modified by the
ligand-bound receptor; the quantity $\dot{n}/p$ is thus the sampling
rate of the receptor, be it ligand bound or not. Multiplied with the
relaxation rate $\tr$, it yields the total number of receptor samples
$\mN$ obtained during $\tr$. However, not all these samples are
reliable. The effective number of samples is $\Neff = q \mN$, where
$0<q<1$ quantifies the quality of the sample. Here, $\beta = 1 /
(k_{\rm B}T)$ is the inverse temperature, $\Delta \mu_1$ and $\Delta
\mu_2$ are the free-energy drops over the activation and deactivation
reaction, respectively, with $\Delta \mu=\Delta \mu_1 + \Delta \mu_2$
the total drop, determined by the fuel turnover (see \fref{model}(b)). If
the system is in thermodynamic equilibrium, $\Delta \mu_1 = \Delta
\mu_2 = \Delta \mu=0$, $q\to 0$ and the system cannot sense, while if
the system is strongly driven and $\Delta \mu_1,\Delta \mu_2 \to
\infty$, $q\to 1$ and $\Neff \to \mN$. Yet, even when all samples are
reliable, they may contain redundant information on the receptor
state. The factor $f_{\rm I}$ is the fraction of the $\Neff$ samples
that are independent. It reaches unity when the receptor sampling
interval $\Delta=2\tr/(\Neff/\RT)$ becomes larger than the receptor
correlation time $\tc$.

When the number of samples becomes very large and $\NI, \Neff \to
\infty$, the sampling error reduces to zero. However, the sensing
error still contains a second contribution, which, following
Ref. \cite{Bowsher2013}, we call the dynamical error. This
contribution only depends on timescales. It arises from the fact that
the samples encode the receptor history and hence the ligand
concentration over the past $\tr$, which will, in general, deviate
from the quantity that the cell aims to predict---the current
concentration $\LT$. Indeed, this contribution yields a systematic
error, which cannot be eliminated by increasing the number of receptor
samples, their independence or their accuracy. It can only be reduced
to zero by making the integration time $\tr$ much smaller than the
ligand timescale $\tL$ (assuming that $\tc$ is typically much smaller
than $\tr,\tL$). Only in this regime will the ligand concentration in
the past $\tr$ be similar to the current concentration, and can the
latter be reliably inferred from the occupancy of the receptor
provided the latter has been estimated accurately by taking enough
samples.

Importantly, the dynamics of the input signal not only affects the
sensing precision via the dynamical error, but also via the sampling
error. This effect is contained in the prefactor of the sampling
error, $(1+\tc/\tL)^2(1+\tr/\tL)^2$, which has its origin in the
dynamic gain $\dg$ (\eref{dg_main}). It determines how the sampling error
$\sigma^{2, \rm samp}_\phL$ in the estimate of $p_\tr$
(\eref{SampErrp_main}) propagates to
the error in the estimate of  $\LT$ (see \eref{SNRp}). Only when
$\tc,\tr\ll \tL$ can the readout system closely track the input
signal, and does $\dg$ reach its maximal value, the static gain $\g$,
thus minimizing the error propagation from $p_\tr$ to $\LT$.

\section{\label{sec:resources} Fundamental resources}

We can use \eref{minfo_final_sampl} to identify the
\textit{fundamental} resources for cell sensing \cite{Govern2014}. A
fundamental resource is a (collective) variable $Q_i$ that, when fixed
to a constant, puts a non-zero lower bound on $\text{SNR}^{-1}$, no
matter how the other variables are varied. It is thus mathematically
defined as:
\begin{align}
{\rm MIN}_{Q_i={\rm const}}\left(\text{SNR}^{-1}\right)=f({\rm const})>0.
\elabel{fundres}
\end{align}
To find these collective variables, we numerically or analytically
minimized $\text{SNR}^{-1}$, constraining (combinations of) variables
yet optimizing over the other variables.  To this end, it is helpful to rewrite
\eref{minfo_final_sampl} by splitting the first term in between the
square brackets of the sampling error and then grouping one term with
the second term using that $\mN_{\rm eff} = q \mN=q\dot{n}\tr
/p$ (see also section S-V):
\begin{widetext}
\begin{align}
\elabel{minfo_final1}
\text{SNR}^{-1} &= \left(1+\frac{\tc}{\tL}\right)^2
\left(1+\frac{\tr}{\tL}\right)^2  \left[\underbrace{\frac{\left(\overline{\LT}/\sigma_\LT\right)^2}{p(1-p)\RT(1+\tr/\tc)}}_\text{receptor input noise} + \underbrace{\frac{\left(\overline{\LT}/\sigma_\LT\right)^2}{(1-p)^2q\dot{n}\tr}}_\text{coding noise}\right]  +\underbrace{\left(1+\frac{\tc}{\tL}\right) \left(1+\frac{\tr}{\tL}\right) \left(1+\frac{\tc \tr}{\tL(\tc+\tr)}\right) -1}_\text{dynamical error}.
\end{align}
\end{widetext}
The first term in between the square brackets describes the
contribution that comes from the stochasticity in the concentration
measurements at the receptor level. 
The second term in between the square brackets, the coding noise,
describes the error that arises in storing these measurements into the
readout molecules.

\begin{figure*}
  \centering
  \includegraphics[width=\textwidth]{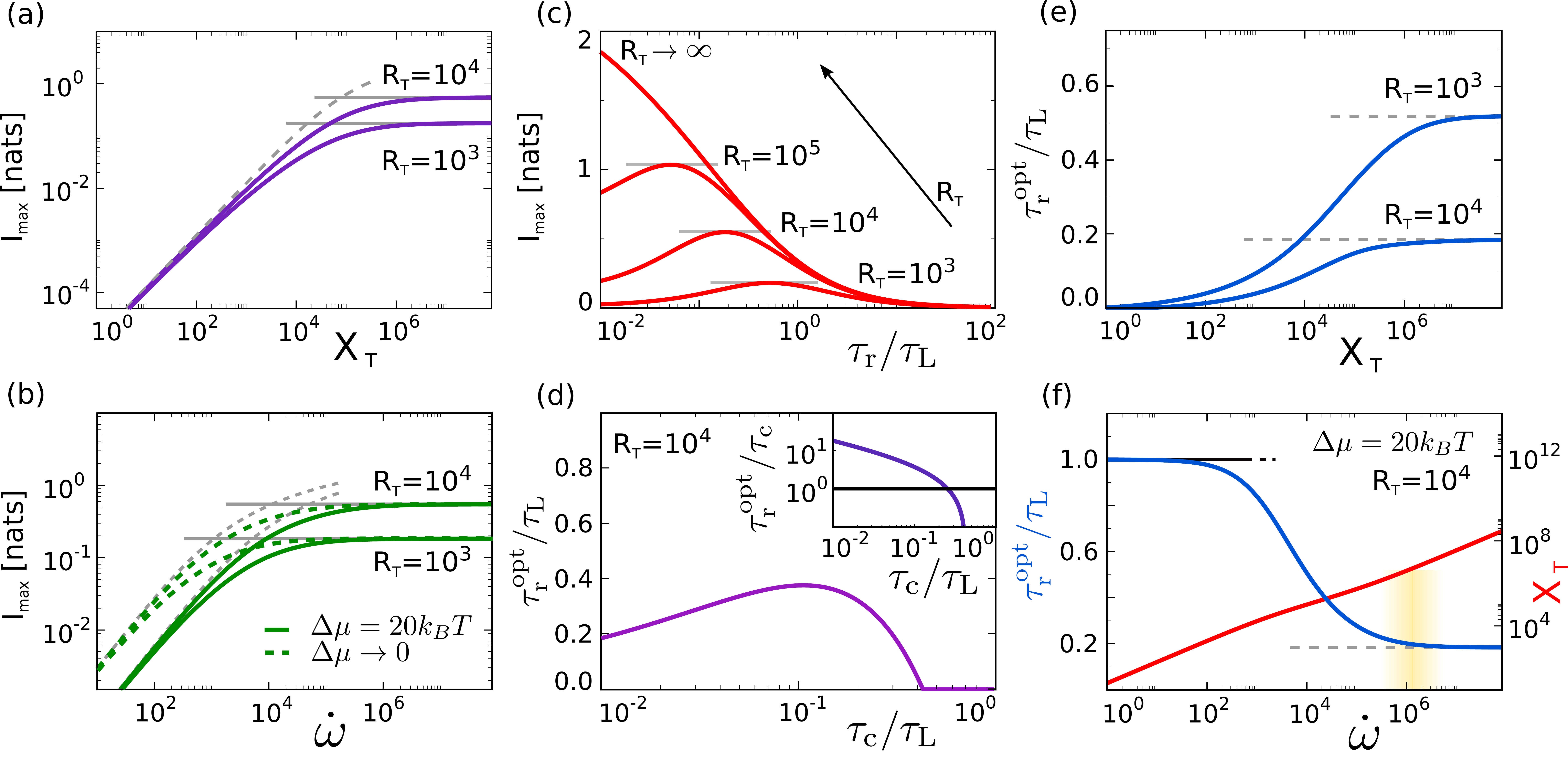}
  \caption{\flabel{main}\textbf{Receptors $\RT$, readout molecules
      $\xT$ and $\dot{w}$ fundamentally limit sensing, and there
      exists an optimal integration time $\tr$ that depends on which
      of the resources is limiting.} (a+b) $\RT$, $\xT$ and $\dot{w}$
    are fundamental resources, with no trade-offs between
    them. Plotted is the maximum mutual information $I_{\rm
        max}=1/2\ln(1+{\rm SNR}_{\rm max})$, obtained by minimizing
    \eref{minfo_final1} over $p$ and $\tr$, for different combinations
    of (a) $\xT$ and $\RT$ in the irreversible limit $q\to 1$, and (b)
    $\dot{w}$ and $\RT$ for two different values of $\Delta \mu$. The
    sensing precision is bounded by the limiting resource, $\RT$
    (solid grey lines, \eref{RTlim2}), $\xT$ (dashed grey line,
    \eref{XTbound}, panel a), or $\dot{w}$ (dashed grey lines,
    \erefstwo{wirr}{wqeq}, panel b).  (c)
    $I_{\rm max}$ as a function of $\tr$ for different values of $\RT$
    in the Berg-Purcell limit ($q\to 1$ and $\xT \to \infty$). There
    exists an optimal integration time $\tr^{\rm opt}$ that maximizes
    the sensing precision; $\tr^{\rm opt}$ decreases with $\RT$. (d)
    In this limit, $\tr^{\rm opt}$ depends non-monotonically on the
    receptor-ligand correlation time $\tc$: it first increases with
    $\tc$ to sustain time-averaging, but then drops when $\tr^{\rm opt}/\tc$
    becomes of order unity
    and time-averaging is no longer effective (see inset). (e) $\tr^{\rm opt}$ as a
    function of $\xT$ for different values of $\RT$. When $\xT < \RT$,
    time averaging is not possible and the optimal system is an
    instantaneous responder, $\tr^{\rm opt}\to 0$; when $\xT \gg \RT$
    the system reaches the Berg-Purcell regime in which $I_{\rm max}$
    is limited by $\RT$ rather than $\xT$ (see panel a). (f) $\tr^{\rm
      opt}$ and $\xT$ as a function of $\dot{w}$. When the
    power $\dot{w} \sim \xT /\tr$ is limiting, the sampling error
    dominates and $\tr^{\rm opt}$ equals $\tL$ to maximize $\xT$,
    minimizing the sampling error; $\tr^{\rm opt}$ then decreases
    to trade part of the decrease in the sampling error for a
    reduction in the dynamical error such that both decrease; when
    the sampling interval $\Delta \sim \tr \RT / \xT$ becomes
    comparable to $\tc$, in the region marked by the yellow bar, the
    sampling error is no longer limited by $\xT$, such that $\tr$ now
    limits both sources of error; the two sources can therefore no
    longer be decreased simultaneously by increasing $\dot{w} \sim \xT
    / \tr$; the system has entered the Berg-Purcell regime where
    $\tr^{\rm opt}$ is determined by $\RT$ rather than $\dot{w}$ (see
    panel (b)). Parameter values unless specified: $\tc / \tL =
    10^{-2}$; $\sigma_L / \overline{L}_T = 10^{-2}$.}
\end{figure*}

\eref{minfo_final1} allows us to identify the fundamental
    resources by constraining combinations of variables while
    optimizing over others by taking limits (see
  \eref{fundres}). As we show below, this reveals that these
     resources are
    the number of receptors $\RT$, their integration time $\tr$, the
    number of readout molecules $\xT$, and the power
    $\dot{w}=\dot{n}\Delta \mu$. \fref{main}
    illustrates that these resources are indeed fundamental, and also
   elucidates the design logic of the system.

Panel (a) of \fref{main} shows the
  maximal mutual information $I_{\rm max}(L;x^*)$ as a function of
  $\xT$ for different values of $\RT$, obtained by optimizing
  \eref{minfo_final1} over $p$ and $\tr$, in the irreversible limit
  $q\to 1$. When $\xT$ is small, $I_{\max}$ cannot be increased by
  raising $\RT$: no matter how many receptors the system has, the
  sensing precision is limited by the pool of readout molecules and
  only increasing this pool can raise $I_{\rm max}$. However, when
  $\xT$ is large, $I_{\rm max}$ becomes independent of $\xT$. In this
  regime, the number of receptors $\RT$ limits the number of
  independent concentration measurements and only increasing $\RT$ can
  raise $I_{\rm max}$. Similarly, panel (b) shows that when the power
  $\dot{w}$ is limiting, $I_{\rm max}$ cannot be increased by $\RT$
  but only by increasing $\dot{w}$. Clearly, the resources receptors,
  readout molecules and energy cannot compensate each other: the
  sensing precision is bounded by the limiting resource. 

Importantly, however, while for sensing static concentrations the
products $\RT \tr / \tc$ (receptors and their integration time) and
$\dot{w} \tr$ (the energy) are fundamental \cite{Govern2014}, for
time-varying signals $\RT$, $\dot{w}$, and $\tr$ separately limit
sensing. Consequently, neither receptors $\RT$ nor power $\dot{w}$ can
be traded freely against time $\tr$ to reach a desired sensing
precision, as is possible for static signals.  There exists an optimal
integration time $\tr^{\rm opt}$ that maximizes the sensing precision,
and its value depends on which of the resources $\RT$, $\xT$ and
$\dot{w}$ is limiting (\fref{main}(c)-(f)). We now discuss these three
regimes in turn.

\subsection{\label{sec:RT} The number of receptors $\RT$} 
As Berg and Purcell pointed out, cells can reduce the sensing error by
increasing the number of receptors or by taking more measurements per
receptor, via the mechanism of time integration \cite{Berg:1977bp}.
In the {\em Berg-Purcell} regime where the receptors and their
integration time are limiting, the coding noise is zero and
\eref{minfo_final1} reduces to
\begin{align}
 \text{SNR}^{-1}
  &\ge \left(1+\frac{\tr}{\tL}\right)^2
  \frac{4\left(\overline{\LT}/\sigma_\LT\right)^2}{\RT \; \tr/\tc} +
  \frac{\tr}{\tL}.\elabel{RTlim2}
\end{align}
This result corresponds to the limits $\xT \to \infty$ and
$\dot{w}\to \infty$ in panels (a) and (b) of \fref{main},
respectively.

\eref{RTlim2} shows that the sensing precision does not depend on $\RT
\tr / \tc$, as for static signals \cite{Govern2014}, but on $\RT$ and
$\tr$ separately, such that an optimal integration time $\tr^{\rm
  opt}$ emerges that maximizes the sensing precision (see
\fref{main}c).  Increasing $\tr$ improves the mechanism of time
integration by increasing the number of independent samples per
receptor, $\tr/\tc$, thus reducing the sampling error
(\eref{minfo_final_sampl}).  However, increasing $\tr$ raises the
dynamical error. Moreover, it lowers the dynamical gain $\dg$, which
increases the propagation of the error in the estimate of the receptor
occupancy to that of the ligand concentration.  The optimal
integration time $\tr^{\rm opt}$ arises as a trade-off between these
three factors.

Fig.~\ref{fig:main}(c) also shows that the optimal integration time
$\tr^{\rm opt}$ decreases with the number of receptors $\RT$. The
total number of independent concentration measurements is the number
of independent measurements per receptor, $\tr/\tc$, times the number
$\RT$ of receptors, $\mN_{\rm I}=\RT \tr/\tc$. As $\RT$ increases, less
measurements $\tr/\tc$ per receptor have to be taken to remove the
receptor-ligand binding noise, explaining why $\tr^{\rm opt}$
decreases as $\RT$ increases. Indeed, the sensing error reduces to
zero when $\RT\to \infty$ and $\tr \to 0$, allowing for optimal signal
tracking.

Interestingly, $\tr^{\rm opt}$ depends non-monotonically on the
receptor-ligand correlation time $\tc$ (Fig.~\ref{fig:main}d).  When
$\tc$ increases at fixed $\tr$, the receptor samples become more
correlated. To keep the mechanism of time integration effective,
$\tr$ must increase as $\tc$ rises. Increasing $\tr$ will, however,
also distort the signal, and to avoid too strong signal distortion the
cell compromises on time integration by decreasing the {\em
  ratio} $\tr/\tc$ (see inset).  When $\tr$ becomes too large, the
benefit of time integration no
longer pays off the cost of signal distortion. Now not only the ratio
$\tr/\tc$ decreases (inset of Fig.~\ref{fig:main}(d)), but also $\tr$
itself (Fig.~\ref{fig:main}(c)). The sensing system switches to a
different strategy. It no longer employs time
integration, but rather becomes an instantaneous responder of the
ligand-binding state of the receptor.

\subsection{\label{sec:XT}The number of readout molecules $X_{\rm T}$}
To implement time integration, the cell needs to store the 
receptor states in the readout molecules.  When the number of readout
molecules $\xT$ is limiting, the coding noise in \eref{minfo_final1}
dominates over the receptor input noise. Noting that the flux
$\dot{n}=f(1-f) q \xT /\tr$, with $f={\overline{x}^* / X_{\rm T}}$ the
fraction of modified readout molecules, we find that in the
irreversible regime ($q\to 1$), the sensing error is bounded by
\begin{align}
\text{SNR}^{-1}
&\ge \left(1+\frac{\tr}{\tL}\right)^2  \left[ \frac{4\left(\overline{\LT}/\sigma_\LT\right)^2}{X_{\rm T}}\right]+\frac{\tr}{\tL}\elabel{XTbound}\\
&\ge \frac{4\left(\overline{\LT}/\sigma_\LT\right)^2}{X_{\rm T}}. 
\elabel{XTlim}
\end{align}
Clearly, $\xT$ is a fundamental resource that puts a hard bound on the
mutual information  (\fref{main}(a)).

\erefstwo{XTbound}{XTlim} show that to reach the sensing limit set by
$\xT$, the receptor integration time $\tr$ needs to be zero. This is
in marked contrast to the non-zero optimal integration $\tr^{\rm opt}$
in the Berg-Purcell regime where $\RT$ is limiting (see
\fref{main}(c)). To elucidate this, \fref{main}(e) shows the optimal
integration time $\tr^{\rm opt}$ as a function of $\xT$.  When $\xT$
is smaller than $\RT$, the average number of samples per receptor is
less than unity. At any given time, there are many receptors whose
concentration measurements are not stored in the downstream readout
molecules. In this regime, the system cannot time integrate the
receptor, and to minimize signal distortion the optimal integration
time $\tr^{\rm opt}$ is essentially zero. However, when $\xT$ is
increased, the likelihood that two or more readout molecules provide a
sample of the same receptor molecule rises, and time averaging becomes
possible. Yet to obtain receptor samples that are independent, the
integration time $\tr$ must be increased to make the sampling interval
$\Delta \sim \tr \RT / \xT$ larger than the receptor correlation time
$\tc$.  As $\xT$ and hence the total number of samples $\mN$ are
increased further, the number of samples that are independent,
$\mN_{\rm I}$, only continues to rise when $\tr$ increases with
$\xT$ further. However, while this reduces the sampling error, it does also
increase the dynamical error. When the decrease in the sampling error
no longer outweighs the increase in the dynamical error, $\tr^{\rm
  opt}$ and the mutual information no longer change with $\xT$ (see
\fref{main}(a)). The system has entered the Berg-Purcell regime in
which $\tr^{\rm opt}$ and the mutual information are given by the
optimization of \eref{RTlim2} (grey dashed line). In this regime,
increasing $\xT$ merely adds redundant samples: the number of
independent samples remains $\mN_{\rm I} = \RT \tr^{\rm opt} / \tc$.

\subsection{\label{sec:power}The power $\dot{w}=\dot{n}\Delta\mu$}
Time integration relies on copying the ligand-binding state of the
receptor into the chemical modification states of the readout
molecules \cite{Mehta2012,Govern2014}. This copy process correlates
the state of the receptor with that of the readout, which requires
work input \cite{Ouldridge:2017hs}. 

The free-energy $\Delta \mu$ provided by the fuel
turnover drives the readout around the cycle of modification and
demodification (\fref{model}). The rate at which the fuel molecules do
work is the power $\dot{w} = \dot{n}\Delta \mu$ and the total work
performed during the integration time $\tr$ is $w\equiv \dot{w}
\tr$. This work is spent on taking samples of receptor molecules that
are bound to ligand, because only they can modify the readout. The
total number of effective samples of ligand-bound receptors during
$\tr$ is $p\Neff$ (\eref{NI}), which means that the work per
effective sample of a ligand-bound receptor is $w/(p\Neff)=\Delta \mu
/ q$ \cite{Govern2014}. 

To understand how energy limits the sensing precision, we can
distinguish between two limiting regimes \cite{Govern2014}.  When
$\Delta \mu > 4 k_{\rm B}T$, the quality factor $q\to 1$ (\eref{NI})
and the work per sample of a ligand-bound receptor is simply
$w/(p\Neff) = \Delta \mu$ \cite{Govern2014}. In this irreversible
regime, the power limits the sensing accuracy not because it limits
the reliability of each sample, but because it limits the rate
$\dot{n}=\dot{w}/\Delta \mu$ at which the receptor is sampled:
\begin{align}
\elabel{wirr}
\text{SNR}^{-1} 
&\ge \left(1+\frac{\tr}{\tL}\right)^2 \left(\frac{\Delta \mu\left(\overline{\LT}/\sigma_\LT\right)^2}{\dot{w}\tr}\right)+\frac{\tr}{\tL},
\end{align}
 obtained from \eref{minfo_final1} by taking $\RT\to \infty$, $p\to 0$.
This expression shows that the sensing precision is fundamentally
bounded not by the work $w=\dot{w} \tr$, as observed for static
signals \cite{Govern2014}, but rather by the power $\dot{w}$ and the
integration time $\tr$ separately such that an optimal integration
time $\tr^{\rm opt}$ emerges (\fref{main}(f)).

When $\Delta \mu < 4 k_{\rm B}T$, the system enters the
quasi-equilibrium regime in which the quality factor $q\to \beta
\Delta \mu / 4$ (see \eref{NI}, noting that in the optimal system
$\Delta \mu_1 = \Delta \mu_2 = \Delta \mu / 2$) \cite{Govern2014}. The
bound on the sensing error (\eref{minfo_final1}) set by the power
constraint now becomes
\begin{align} \text{SNR}^{-1}
&\ge \left(1+\frac{\tr}{\tL}\right)^2
  \left(\frac{4k_{\rm B}T
      \left(\overline{\LT}/\sigma_\LT\right)^2}{\dot{w}\tr}\right)+\frac{\tr}{\tL}\elabel{wqeq}.
\end{align}
Comparing this expression to \eref{wirr}, which only holds when
$\Delta \mu > 4 k_{\rm B}T$, it is clear that the sensing error is
minimized in the quasi-equilibrium regime, see \fref{main}(b). This
regime maximizes the number of effective measurements per work input,
because the work per effective measurement reaches its fundamental
lower bound, $w/(p\Neff)=\Delta \mu/q=4 k_{\rm B}T$ \cite{Govern2014}.

While the sensing precision for a given power and time constraint is
higher in the quasi-reversible regime, more readout molecules are
required to store the concentration measurements in this
regime. Noting that the flux $\dot{n} = f(1-f) \xT q / \tr =
\dot{w}/\Delta \mu$ (Eq. S114), it follows that in the irreversible
regime ($q\to 1$) the number of readout molecules consuming energy at
a rate $\dot{w}$ is
\begin{align}
\xT^{\rm irr} = \frac{\dot{w}\tr}{\Delta \mu f(1-f)}
\elabel{Xirr}
\end{align}
 while in the quasi-equilibrium regime
($q\to\Delta \mu/4$) it is 
\begin{align}
\xT^{\rm qeq} = \frac{\dot{w}\tr 4k_{\rm B}T}{\Delta \mu^2
f(1-f)}.
\elabel{Xqeq}
\end{align}
Since in the quasi-equilibrium regime $\Delta \mu < 4 k_{\rm B}T$,
$\xT^{\rm qeq}>\xT^{\rm irr}$. 

\fref{main}(f) shows how the optimal integration time $\tr^{\rm opt}$
depends on the power $\dot{w}$.  Since the system cannot sense without
any readout molecules, in the low power regime the system maximizes
$\xT$ subject to the power constraint $\dot{w} \sim \xT /\tr$ (see
\erefstwo{Xirr}{Xqeq}) by making $\tr$ as large as possible, which is
the signal correlation time $\tL$---increasing $\tr^{\rm opt}$ further
would average out the signal itself. As $\dot{w}$ is increased, $\xT$
rises and the sampling error decreases. When the sampling error
becomes comparable to the dynamical error (\eref{minfo_final_sampl}),
the system starts to trade a further reduction in the sampling error
for a reduction in the dynamical error: $\tr^{\rm opt}$ now goes
down. In this regime, the sampling error and the dynamical error are
reduced simultaneously by increasing $\xT$ and decreasing $\tr^{\rm
  opt}$. This continues until the sampling interval $\Delta \sim \RT
\tr / \xT$ becomes comparable to the receptor correlation time $\tc$,
as marked by the yellow bar. Beyond this point, $\Delta < \tc$ and the
sampling error is no longer limited by $\xT$ but rather by $\tr$,
since $\tr$ bounds the number of independent samples per
receptor, $\tr/\tc$. Because $\tr$ now limits both sources of
error, they can no longer be reduced simultaneously. The
system has entered the Berg-Purcell regime, where $\tr^{\rm opt}$ is
determined by the trade-off between the dynamical error and the
sampling error as set by the maximum number of independent samples,
$\RT\tr / \tc$ (\fref{main}(c)).

\section{\label{sec:alloc}The optimal allocation principle, revisited}
In sensing static concentrations, there exists three fundamental
classes of resources: receptors and their integration time $\RT
\tr/\tc$, readout molecules $\xT$, and energy $\dot{w}\tr$ injected
during $\tr$ \cite{Govern2014}. These fundamental resource classes
cannot compensate each other in achieving a desired sensing
precision---they limit sensing like weak links in a chain. It means
that in an optimally designed system each class is equally limiting so
that no resource is wasted.  This yields the design principle that in
an optimal system $\RT \tr / \tc \approx \xT \approx \beta \dot{w}\tr$
\cite{Govern2014}. However, in sensing time-varying signals, a
trade-off between time integration and signal tracking is inevitable.
As a result, besides $\xT$, the receptors $\RT$, the power $\dot{w}$
and the integration time $\tr$ are each fundamental.

Can we nonetheless formulate a similar design principle? We
cannot simply equate the bounds set by the number of receptors $\RT$ and their
integration time $\tr$ (\eref{RTlim2}), the number of readout molecules
$\xT$ (\eref{XTlim}) and the power $\dot{w}$ (\eref{wqeq}),
because they correspond to different sensing strategies: when $\RT$
is limiting, there exists an
optimal non-zero integration time $\tr^{\rm opt}$, while if $\xT$ is
limiting $\tr^{\rm opt}\approx 0$, as discussed above. 

Remarkably, however, \erefsthree{RTlim2}{XTbound}{wqeq} have the
  same functional form $f(x)$, with $x=\RT\tr/\tc, \xT, \beta
  \dot{w}\tr$, respectively. This means that when {\em for a given $\tr$},
  $\RT \tr / \tc = \xT = \beta \dot{w}\tr$ and $f(\RT
  \tr/\tc)=f(\xT)=f(\beta\dot{w}\tr)$, the bounds on the sensing precision
  as set by, respectively, the number of receptors $\RT$
  (\eref{RTlim2}), the number of readout molecules $\xT$
  (\eref{XTbound}), and the power $\dot{w}$ (\eref{wqeq}), are equal.
  Each of these resources is now equally limiting sensing and no
  resource is in excess. We thus recover the optimal resource
  allocation principle originally formulated for systems sensing
  static concentrations \cite{Govern2014}:
\begin{align}
\RT \tr/\tc \approx
  \xT \approx \beta \dot{w} \tr.
  \elabel{optAllocOrg}
\end{align}
Irrespective of whether the concentration fluctuates in time, the
number of independent concentration measurements at the receptor level
is $\RT \tr / \tc$, which in an optimally designed system also equals the
number of readout molecules $\xT$ and the energy $\beta \dot{w}\tr$ that are
both necessary and sufficient to store these measurements reliably.

Importantly, \eref{optAllocOrg} holds for any integration time $\tr$,
yet it does not specify $\tr$. What is the optimal $\tr$ that
minimizes the sensing error?  The design principle $\RT\tr/\tc=\xT$
means that for a fixed $\xT$, $\RT$ can be increased by simultaneously
decreasing $\tr$. This increases the sensing precision (see
  Fig. S1, S-VII). In fact, for a fixed $\xT$, the precision is
maximized when $\RT=\xT$ and $\tr=0$, because in this limit the
dynamical error is zero. However,  the power diverges in
this limit, because in the optimal system $\beta
\dot{w}\tr\approx \xT$ (\eref{optAllocOrg}).

Intriguingly, the cell membrane is highly crowded and many systems
employ time integration \cite{Berg:1977bp,Bialek2005,Govern2014}. This
suggests that these systems employ time integration and accept the
signal distortion that comes with it, simply because there is not
enough space on the membrane to increase $\RT$. Our theory then
  allows us to predict
the optimal integration time $\tr^{\rm opt}$ based on the premise
that $\RT$ is limiting.  As \eref{RTlim2} reveals, in this limit
$\tr^{\rm opt}$ does not only depend on $\RT$, but also on $\tc$,
$\tL$, and $\sigma_\LT/\overline{\LT}$: $\tr^{\rm opt}=\tr^{\rm
  opt}(\RT,\tc,\tL,\sigma_{\LT}/\overline{\LT})$. The optimal design
of the system is then given by \eref{optAllocOrg} but with $\tr$ given
by $\tr^{\rm opt} = \tr^{\rm
  opt}(\RT,\tc,\tL,\sigma_{\LT}/\overline{\LT})$:
\begin{align}
  \RT \tr^{\rm opt}/\tc \approx
  \xT^{\rm opt} \approx \beta \dot{w}^{\rm opt} \tr^{\rm opt}.
  \elabel{optAlloc}
\end{align}
This design principle maximizes for a given number of receptors $\RT$
the sensing precision, and minimizes the number of readout molecules
$\xT$ and power $\dot{w}$ needed to reach that precision.

\section{\label{sec:taurexp}Comparison with experiments}

If the number of receptors is limiting the sensing precision, then
  our theory predicts an optimal integration time $\tr^{\rm
    opt}(\RT,\tc,\tL,\sigma_{\LT}/\overline{\LT})$ that is given by
  \eref{RTlim2}.  We can test this prediction for the chemotaxis
  system of the bacterium {\it E. coli}, which has been well
  characterized experimentally. In this system, the receptor forms a
  complex with the kinase CheA. This complex, which can be
  coarse-grained into $R$ \cite{Govern2014}, can bind the ligand L and
  activate the intracellular messenger protein CheY ($x$) by
  phosphorylating it. Deactivation of CheY is catalyzed by CheZ, the
  effect of which can be coarse-grained into the deactivation
  rate. The {\it E. coli} chemotaxis system also exhibits adaptation
  on longer timescales, due to receptor methylation and
  demethylation. However, the integration time for the receptor-ligand
  binding noise is not given by the adaptation timescale, but rather
  by the relaxation rate of the push-pull network that controls CheY
  (de) phosphorylation
  \cite{Sartori:2011fh}.
  
\begin{figure*}
  \centering
  \includegraphics[width=0.75\textwidth]{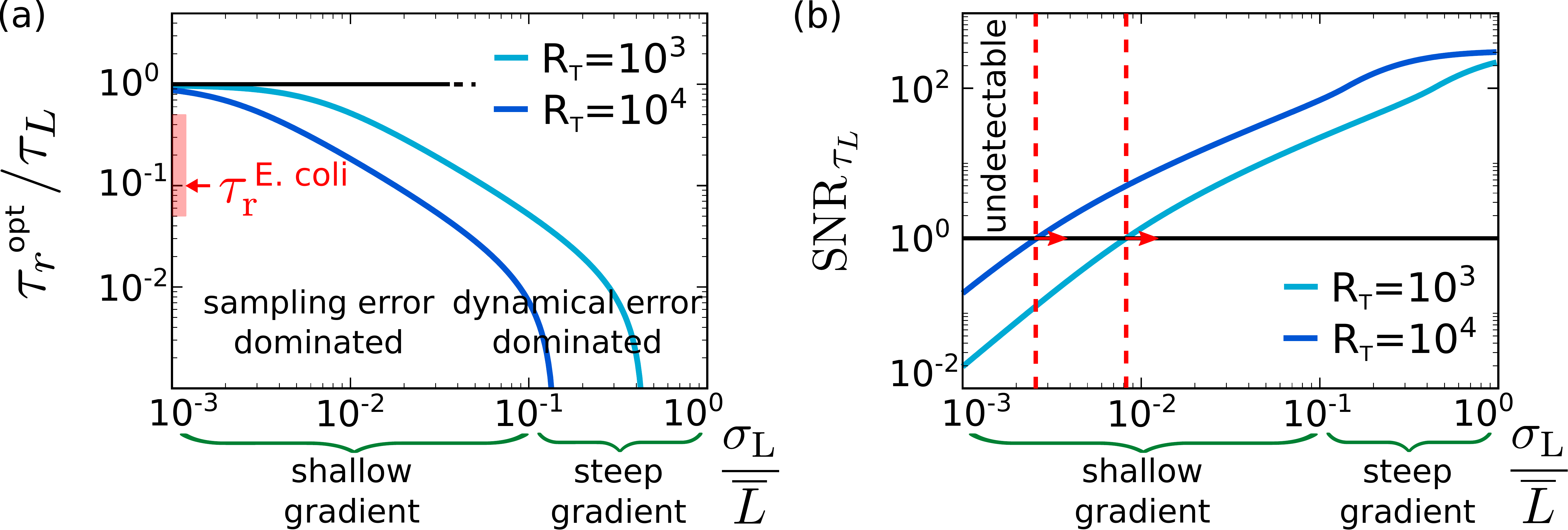}
  \caption{\textbf{The optimal integration time for the chemotaxis
      system of {\it E. coli}}. (a) The optimal integration time
    $\tr^{\rm opt}$, obtained by numerically optimizing \eref{RTlim2},
    as a function of the relative strength of the input noise,
    $\sigma_\LT/\overline{\LT}$, for two different copy numbers $\RT$
    of the receptor-CheA complexes; for an exponential gradient with
    length scale $x_0$, the relative noise strength
    $\sigma_\LT/\overline{\LT} \simeq l /x_0$, where $l\approx 50
    \mu{\rm m}$ is the run length of {\it E. coli}. It is seen that
    $\tr^{\rm opt}$ increases as $\sigma_\LT/\overline{\LT}$
    decreases. This is because the relative importance of the sampling
    error compared to the dynamical error increases, necessitating a
    longer integration time. The figure also shows that $\tr^{\rm
      opt}$ decreases as $\RT$ is increased, because that allows for
    more instantaneous measurements (see also \fref{main}). The red
    bar indicates the range of the estimated integration time of {\it
      E. coli}, $0.05 {\rm ms} < \tr < 0.5 {\rm ms}$, based on its
    attractant and repellent response respectively
    \cite{Sourjik2002}, divided by the input timescale $\tL \approx
    1 {\rm s}$ based on its typical run time of about a second
    \cite{Berg:1972wt,Taute:2015ce}. The panel indicates that {\it
      E. coli} has been optimized to detect shallow concentration
    gradients. (b) The \snr\ $\text{SNR}_\tL=(\sigma_L/\delta
    \hat{\LT})^2\tL/\tr$ as a function of $\sigma_\LT/\overline{\LT}\simeq
    l/x_0$. To be able to detect the gradient, the $\text{SNR}_\tL$ must
    exceed unity. The panel shows that the shallowest gradient that
    {\it E. coli} can detect (marked with dashed red line) has, for
    $\RT=10^4$, a length scale of $x_0\approx 25000 \mu{\rm m}$
    (corresponding to $\sigma_L/ \overline{\LT}\approx2\times 10^{-3}$),
    which is consistent with experiments based on ramp responses
    \cite{Shimizu:2010ig}. Other parameter: receptor-ligand binding
    correlation time $\tc=10{\rm ms}$
    \cite{Vaknin2007,Danielson1994}.}
  \flabel{Fig_tau_r}
\end{figure*}

To test the prediction  for $\tr^{\rm
  opt}(\RT,\tc,\tL,\sigma_{\LT}/\overline{\LT})$, we need to estimate
$\RT$, $\tc$, $\tL$ and $\sigma_{\LT}/\overline{\LT}$.
The number of receptor-CheA complexes depends on the growth rate and
varies between $\RT=10^3$ and $\RT=10^4$ \cite{Li:2004eh}. The dissociation
constant for the binding of aspartate to the Tar receptor is $K_D
\approx 0.1 \mu{\rm M}$ \cite{Vaknin2007}, which with an association
rate of $k_{\rm on} \approx 10^9{\rm M}^{-1}/{\rm s}$
\cite{Danielson1994} yields a receptor-ligand dissociation rate of
$k_{\rm off} \approx 100 {\rm s}^{-1}$. Protein occupancies are
  typically in the range $0.1-1$ and following our previous work we
  assume $p=0.5$ \cite{Govern2014}, which 
gives a receptor-ligand correlation time of $\tc \simeq 1 / (2 k_{\rm
  off}) \approx 10 {\rm ms}$. The timescale $\tL$ of the input
fluctuations is set by the typical run time, which is on the order of
a few seconds, $\tL \approx 1 {\rm s}$
\cite{Berg:1972wt,Taute:2015ce}.

This leaves one important parameter to be determined, the relative
variance of the ligand concentration fluctuations, $(\sigma_\LT /
\overline{\LT})^2$. This is set by the spatial ligand-concentration
  profile and by the typical length of a run. We have a good estimate
  of the latter; in shallow gradients it is on the order of $l\simeq
  50 \mu{\rm m}$
  \cite{Berg:1972wt,Taute:2015ce,Jiang:2010gja,Flores:2012is}. However,
  we do not know the spatial concentration profiles that {\it E. coli}
  has experienced during its evolution.  For this reason we will study
  the optimal integration time as a function of $(\sigma_\LT /
  \overline{\LT})^2$. We can however get a sense of the scale by
  considering an exponential ligand-concentration gradient. For a
  profile $\overline\LT (x) = \LT_0 e^{x/x_0}$ with length scale
  $x_0$, the relative change in the signal over the length of a run is
  $\sigma_\LT / \overline{\LT} \simeq (d\overline{\LT}/dx) l /
  \overline{\LT}= l/x_0$.  Experiments indicate that for $x_0 \gtrsim
  500\mu{\rm m}$ the cells reach a stable drift velocity before the
  receptor saturates \cite{Shimizu:2010ig,Flores:2012is}. Inspired by
  these observations, we consider the range $\sigma_\LT /
  \overline{\LT} \approx l / x_0 < 1$, where
  $\sigma_{\LT}/\overline{\LT} < 0.1$ corresponds to shallow
  gradients with $x_0 \gtrsim 500 \mu{\rm m}$ in which the cells move
  with a constant speed \cite{Shimizu:2010ig,Flores:2012is}.

\fref{Fig_tau_r} shows the result. Panel (a) shows that as the
gradient becomes steeper and  $\sigma_\LT / \overline{\LT}\approx l / x_0$
increases, the optimal integration time $\tr^{\rm opt}$ decreases,
dropping to zero when $\sigma_\LT / \overline{\LT} > 0.1$. We can
understand the qualitative behavior by noting that the relative
importance of the dynamical error as compared to the sampling error
scales with $\left(\sigma_\LT / \overline{\LT}\right)^2$ (see
\eref{RTlim2}). Hence, when the gradient is shallow and $\sigma_\LT /
\overline{\LT}$ is small, the dynamical error is small compared to the
sampling error, which allows for a larger optimal integration time
$\tr^{\rm opt}$; at the same time, our theory predicts that $\tr^{\rm
  opt}$ depends on the input timescale $\tL$, such that even in very
shallow gradients $\tr^{\rm opt}$ is bounded by $\tL$. In contrast, in
steep gradients, $\sigma_\LT / \overline{\LT}$ and hence the dynamical
error will be large, which necessitates a small $\tr^{\rm opt}$. In
fact, for $\sigma_\LT / \overline{\LT}>0.1$, the optimal system is an
instantaneous responder.

Experiments indicate that the relaxation rate of CheY is
$\tr^{-1}\approx 2 {\rm s}^{-1}$ for the attractant response and
$\approx 20 {\rm s}^{-1}$ for the repellent response
\cite{Sourjik2002}, such that the integration time $\tr \approx 50 -
500 {\rm ms}$ \cite{Sourjik2002,Govern2014}. \fref{Fig_tau_r}(a) shows
that, according to our theory, this integration time is optimal for
detecting shallow gradients, in the range $l / x_0 \approx \sigma_\LT
/ \overline{\LT} \approx 10^{-3} - 10^{-1}$. Our theory thus suggests
that the sensing system of {\it E. coli} has been optimized for
sensing shallow gradients.

While \fref{Fig_tau_r} indicates that the sensing system of {\it
    E. coli} has been optimized for detecting shallow gradients, it
  does not tell us whether cells can actually do so. To navigate,
  the cells must be able to resolve the signal change over a run. This
  means that the signal-to-noise ratio ${\rm SNR}_\tL$ for the
  concentration measurements during a run of duration $\tL$ must at
  least be of order unity. If the ${\rm SNR}_\tL$ is close to unity,
  it indicates that the system operates close to its fundamental
  sensing limits.

The signal change over a run is $\sigma^2_\tL$ and the effective
  error $(\delta \hat{L})^2 / (\tL / \tr)$ on the concentration
  measurements during a run is the instantaneous sensing error
  $(\delta \hat{L})^2$ divided by the number of independent
  concentration measurement $\tL / \tr$ taken during a run of duration
  $\tL$. The \snr\ for these measurements is thus
  $\text{SNR}_\tL\equiv (\sigma_\LT / \delta \hat{L})^2\tL / \tr$. It
  is plotted in \fref{Fig_tau_r}(b) for the optimized system, with
  $\tr$ equal to the optimal integration time $\tr^{\rm opt}$ that
  maximizes the sensing precision, given by \eref{RTlim2}.

  \fref{Fig_tau_r} shows that our theory predicts that when $\RT=10^3$,
    the shallowest gradient that cells can resolve, defined by
    $\text{SNR}_\tL=1$, is $l / x_0 \approx \sigma_\LT /
    \overline{\LT} \approx 1 \times 10^{-2}$, corresponding to $x_0
    \approx 7500 \mu{\rm m}$, while when $\RT=10^4$, it is $l / x_0 \approx
    2 \times 10^{-3}$ corresponding to $x_0 \approx 25 000 \mu{\rm m}$; the
    shallowest gradient is thus on the order of $x_0\approx
    10^4\mu{\rm m}$. Interestingly, Fig. 2A of \cite{Shimizu:2010ig} shows
    that {\it E. coli} cells can detect exponential up ramps with rate
    $r=0.001/{\rm s}$; using $r=v_{\rm r}/x_0$ where $v_{\rm r}\approx
    10 \mu{\rm m}/{\rm s}$ is the {\it E. coli} run speed
    \cite{Jiang:2010gja}, this means that these cells are indeed able
    to sense very shallow gradients with $x_0 \approx 10^4
    \mu{\rm m}$.
    Importantly, the predictions of our theory, \fref{Fig_tau_r},
    concern the shallowest gradient that the system with the optimal
    integration time can resolve: for any other integration time, the
    shallowest gradient will be steeper.  These observations indicate
    that the optimal integration time is not only sufficient to make
    navigation in shallow gradients possible, but also
    necessary: to enable the detection of shallow gradients with $x_0
    \approx 10^4 \mu{\rm m}$, as observed experimentally
    \cite{Shimizu:2010ig}, the integration time {\em must} have been
    optimized. This is a strong prediction, since it implies that
    evolution has pushed the system to its sensing limits to enable
    navigation in shallow gradients.

\fref{Fig_tau_r} also shows that $\tr^{\rm opt}$ decreases as the
number of receptor-CheA complex, $\RT$, increases. As discussed in
section \ref{sec:RT}, this is because a larger number of receptors
allows for more instantaneous measurements, reducing the need for time
integration. Interestingly, the data of Li and Hazelbauer
\cite{Li:2004eh} shows that the copy numbers of the chemotaxis
proteins vary with the growth rate. Unfortunately, however, the
response time has not been measured as a function of the growth
rate. Clearly, it would be of interest to directly measure the
response time in different strains under different growth conditions.

\section{Discussion}

Here, we have integrated ideas from Refs.
\cite{Tostevin2010,Hilfinger:2011ev,Bowsher2013} on information
transmission via time-varying signals with the sampling framework of
Ref. \cite{Govern2014} to develop a unified theory of cellular
sensing. The theory is founded on the concept of the dynamic
input-output relation $p_\tr(\LT)$. It allows us to develop the idea
that the cell employs the readout system to estimate the average
receptor occupancy $p_\tr$ over the past integration time $\tr$ and
then exploits the mapping $p_\tr(\LT)$ to estimate the current ligand
concentration $\LT$ from $p_\tr$. The error in the estimate of $\LT$ is then
determined by how accurately the cell samples the receptor state to
estimate $p_\tr$, and by how much the ligand concentration in the past
$\tr$, which determines $p_\tr$, reflects the current ligand
concentration. These two distinct sources of error give rise to the
sampling error and the dynamical error in \eref{minfo_final_sampl},
respectively.

While the system contains no less than 11 parameters,
\eref{minfo_final_sampl} provides an intuitive expression for the
sensing error in terms of collective variables that have a clear
interpretation.  The dynamical error is only determined by the input
noise strength $\sigma_L/\overline{L}$ and the timescales in the
problem---the correlation time $\tL$ of the input signal, the receptor
correlation time $\tc$, and the receptor integration time $\tr$. The
sampling error depends on the number of receptor samples, their
independence, and their accuracy---these determine how accurately the
receptor occupancy $p$ is estimated---and on the timescales $\tr, \tc,
\tL$ via the dynamical gain---this determines how the error in $p$
propagates to the estimate of the
concentration. \eref{minfo_final_sampl} shows that even when an
infinite amount of cellular resources is devoted to sensing, reducing
the sampling error to zero, the sensing error is still limited by the
dynamical error when the integration time $\tr$ is finite. The
dynamical error is a systematic error, which can only be eliminated by
reducing $\tr$ to zero. However, while this increases the dynamic
gain (which helps to reduce the sampling error), decreasing $\tr$
ultimately raises the sampling error, because the maximum number of
independent concentration measurements per receptor is bounded by
$\tr$. \eref{minfo_final_sampl} thus predicts that there exists an
optimal integration time that optimizes the trade-off between
minimizing the sampling error and the dynamical error.

Our study reveals that the optimal integration time $\tr^{\rm opt}$
depends in a non-trivial manner on the design of the system. When the
number of readout molecules $\xT$ is smaller than the number of
receptors $\RT$, time integration is not possible and the optimal
system is an instantaneous responder with $\tr^{\rm opt}=0$.  When the
power $\dot{w} \sim \xT / \tr$, rather than $\xT$, is limiting,
$\tr^{\rm opt}$ is determined by the trade-off between the sampling
error, set by $\xT$, and by the dynamical error, set by $\tr$. In both
scenarios, however, one resource, $\xT$ or $\dot{w}$, is limiting the
sensing precision. The other resources do not contribute to
  reducing the sensing error and are thus in excess, making these
  systems suboptimal.

In an optimally designed system all resources are equally limiting
  so that no resource is wasted. This yields the resource allocation
  principle, \eref{optAllocOrg}, first identified in
  Ref. \cite{Govern2014}. That this design principle can be
  generalized to time-varying signals is not obvious because the
  sensing limits associated with the fundamental resources $\RT$,
  $\xT$, and $\dot{w}$, are different, corresponding to different
  sensing strategies with different $\tr^{\rm opt}$. However, our
  theory explains why \eref{optAllocOrg} nonetheless still holds.  The
  dynamics of the input signal affects both the dynamical error and
  the sampling error (see \eref{minfo_final_sampl}), but it influences
  the latter only via the dynamic gain, which influences how the error
  in the estimate of $p_\tr$ propagates to that in $\LT$. The input
  dynamics does not affect the error $\sigma^{2,{\rm
      samp}}_\phL$ in estimating $p_\tr$ itself. Conversely, while
  $\sigma^{2,{\rm samp}}_\phL$ depends on $\RT$, $\xT$ and $\dot{w}$
  since they determine how accurately the receptor is sampled, the
  dynamical error and the dynamic gain do not depend on these
  resources but only on timescales. It is this non-trivial
  decomposition of the sensing error, which explains why the sensing
  limits set by the respective resources for a given $\tr$
  (\erefsthree{RTlim2}{XTlim}{wqeq}) have the same functional form,
  and why the allocation principle can be generalized.  The design
  principle concerns the optimal allocation of resources for
  estimating $p_\tr$, and this holds for any type of input signal: the
  number of independent concentration measurements at the receptor
  level is $\RT\tr/\tc$, irrespective of how the input varies, and in
  an optimally designed system this also equals the number of readout
  molecules $\xT$ and energy $\beta \dot{w}\tr$ to store these
  measurements reliably.

While the allocation principle \eref{optAllocOrg} holds for any
  $\tr$, it does not specify the optimal integration time. However,
  our theory predicts that if the number of receptors is limiting,
  then there exists an optimal integration time $\tr^{\rm opt}$ that
  maximizes the sensing precision for that number of receptors $\RT$
  (\eref{RTlim2}).  Via the allocation principle \eref{optAlloc},
  $\RT$ and $\tr^{\rm opt}$ then together determine the minimal number
  of readout molecules $\xT$ and power $\dot{w}$ to reach that
  precision.  The resource allocation principle together with the
  optimal integration time thus completely specify the optimal design
  of the sensing system for a given number of receptors. 

Our theory, via \erefstwo{RTlim2}{optAlloc}, illuminates how the
  optimal design of a cellular sensing system depends on the dynamics
  of the input signal. In an optimal system, each receptor is sampled
  once every receptor-ligand correlation time $\tc$, $\Delta \approx
  \tc$, and the number of samples per receptor is $\tr^{\rm opt} /
  \Delta \approx \tr^{\rm opt}/\tc$; the optimal integration
  time $\tr^{\rm opt}$ is determined by the trade-off between the age
  of the samples and the number required for averaging the receptor
  state.  When the input signal varies more rapidly and $\tL$
  decreases, the samples need to be refreshed more regularly; to keep
  the dynamical error and the dynamic gain constant, $\tr^{\rm opt}$
  must decrease linearly with $\tL$, see
  \eref{minfo_final_sampl}. Yet, only decreasing $\tr^{\rm opt}$ would
  inevitably increase the sampling error $\sigma^{2, \rm samp}_\phL$
  in estimating the receptor occupancy, because the sampling interval
  $\Delta \sim \RT \tr^{\rm opt} / \xT^{\rm opt}$ would become smaller
  than $\tc$, causing the samples to contain redundant information. To
  keep the sensing precision constant, the number of receptors $\RT$
  needs to be raised with $\tL^{-1}$, such that the sampling
  interval $\Delta \sim \RT \tr^{\rm opt} / \xT^{\rm opt}$ is again of
  order $\tc$, and the decrease in the number of samples per receptor,
  $\tr^{\rm opt}/\tc$, is precisely compensated for by the increase in
  $\RT$. The total number of independent concentration measurements,
  $\RT \tr^{\rm opt}/\tc$, and hence the number of readout molecules
  $\xT^{\rm opt}$ to store these measurements, does indeed not
  change. In contrast, the required power $\dot{w}^{\rm opt}=
  \dot{n}\Delta \mu \sim \xT^{\rm opt} \Delta \mu/ \tr^{\rm opt}$ does
  increase: the readout molecules sample the receptor at a higher rate
  $\dot{n}\sim \xT^{\rm opt} / \tr^{\rm opt}$. Our theory thus
  predicts that when the input varies more rapidly, the number
  of receptors and the power must rise to maintain a required
  sensing precision, while the number of readout molecules does not.

While our theory makes concrete predictions on the optimal ratios of
$\RT$, $\xT$, $\dot{w}$ and $\tr$ given the statistics of the input
signal, it does not predict what the optimal sensing precision and
hence the absolute magnitudes of these resources are. In principle the
cell can reduce the sensing error arbitrarily by increasing $\RT$ and
decreasing $\tr$. Yet, the resource allocation principle,
\eref{optAlloc}, shows that then not only the number of readout
molecules needs to be raised, but also the power. Clearly, improving
the sensing precision comes at a cost: more copies of the components
of the sensing system need to be synthesized every cell cycle, and
more energy is needed to run the system. The optimal sensing precision
is determined by the trade-off between the fitness benefit of sensing
and the energetic cost of maintaining and running the sensing system,
which is beyond the scope of our theory. We emphasize, however, that
the resource allocation principle, \eref{optAlloc}, by itself is
independent of the cost of the respective resources \cite{Govern2014}:
resources that are in excess cannot improve sensing and are thus
wasted, no matter how cheap they are. It probably explains why our
theory, without any fit parameters, not only predicts the integration
time that allows {\it E. coli} to sense shallow gradients
(\fref{Fig_tau_r}), but also the number of receptor and readout
molecules \cite{Govern2014}.

In our study we have limited ourselves to a canonical push-pull
motif. However, the work of Ref. \cite{Govern2014} indicates that our
results hold more generally, pertaining also to sensing systems that
employ cooperativity, negative or positive feedback, or consist of
  multiple layers, as the MAPK cascade. While multiple layers and
feedback change the response time, they do not make time integration
more efficient in terms of readout molecules or energy
\cite{Govern2014}. And provided it does not increase the
  correlation time of the signal \cite{Skoge2011,tenWolde:2016ih},
  cooperative ligand binding can reduce the sensing error per sample,
  but the resource requirements in terms of readout molecules and
  energy per sample do not change \cite{Govern2014}. In all these
systems, time integration requires that the history of the receptor is
stored, which demands protein copies and energy.
 
Our performance measure---the precision by which the system can
estimate the current concentration---is similar to that used to
quantify the accuracy of measuring static concentrations
\cite{Berg:1977bp,Bialek2005,Wang2007,Rappel2008,Endres2009,Hu2010,Mora2010,Govern2012,Kaizu2014,Govern2014}.
This is the natural measure if one is interested in the question how
accurately a cell can respond to the current concentration. Another
performance measure is the learning rate or information flow, which
quantifies the rate at which the system acquires information about the
concentration
\cite{Barato2014,Horowitz:2014wb,Hartich:2016gs,Brittain:2017hf}. An
interesting question for future work would be whether systems that
optimize the learning rate obey a resource allocation principle.

Lastly, in this paper we have studied the resource requirements for
estimating the current concentration via the mechanism of time
integration.  However, to understand how {\it E. coli} navigates in a
concentration gradient, we do not only have to understand how the
system filters the high-frequency ligand-binding noise via time
averaging, but also how on longer timescales the system adapts to
changes in the ligand concentration \cite{Sartori:2011fh}. This
adaptation system also exhibits a trade-off between accuracy, speed
and power \cite{Lan:2012in,Sartori:2015hs}. Intriguingly, simulations
indicate that the combination of sensing (time integration) and
adaptation allows {\it E. coli} not only to accurately estimate the
current ligand concentration, but also predict the future ligand
concentration \cite{Becker2015}. It will be interesting to see whether
an optimal resource allocation principle can be formulated for systems
that need to predict future ligand concentrations.

\begin{acknowledgments}
  We wish to acknowledge Bela Mulder, Tom Shimizu and Tom Ouldridge for many
  fruitful discussions and a careful reading of the manuscript.  This
  work is part of the research programme of the Netherlands
  Organisation for Scientific Research (NWO) and was performed at the
  research institute AMOLF.
\end{acknowledgments}




\setcounter{equation}{0}
\setcounter{figure}{0}
\setcounter{table}{0}
\setcounter{page}{1}
\setcounter{section}{0}
\makeatletter

\renewcommand{\thefigure}{S\arabic{figure}}
\renewcommand{\theequation}{S\arabic{equation}}
\renewcommand{\thetable}{S\arabic{table}}
\renewcommand{\thesection}{S-\Roman{section}}



\begin{center}
\section*{Supporting Information}
\end{center}

{\bf Overview} In this Supporting Information we derive the \snr\
within the sampling framework, Eq. 20 of the main text, which is the
principal result of our work. In this framework, the cell discretely
samples the receptor state to estimate the average occupancy $p_\tr$
over the past integration time, and then inverts the dynamical
input-output relation $p_\tr(\LT)$ to obtain the estimate for the
current concentration $\LT(t)$.

First, however, we review the system and discuss the chemical Langevin
equations that describe it. Then, in section \ref{app:SNRx}, we derive
the expression for the sensing error based on estimating the
concentration from the number of readout molecules $x^*$,
\eref{minfoSI}. This is Eq. 4 of the main text.

In section \ref{app:samplerrors} we derive the principal result of our
work, the sensing error within the sampling framework, Eq. 20 of
  the main text. In \ref{app:xtoptr} we show that this result, for
estimating the concentration from the time-averaged receptor
occupancy, is the same as Eq. 4 and \eref{minfoSI}, for
estimating the concentration from $x^*$.

In the next section, section \ref{app:11to7}, we show how Eq. 20 of
the main text can be rewritten as Eq. 23 of the main text. In section
\ref{app:dynamicgainOpt} we discuss the optimal integration time while
in \ref{app:OptDes} we provide background information on the optimal
resource allocation principle, Eq. 32 of the main text.

\section{The system}
The signal has a
variance $\sigma^2_\LT$ and is assumed to relax exponentially with a
correlation time $\tL=\lambda^{-1}$, as characterized by the
correlation function $\avg{\delta \LT(t)\delta
  \LT(t^\prime)}=\sigma^2_\LT e^{-\lambda(t-t^\prime)}$. The ligand
can stochastically bind the receptor, while the ligand-bound
receptor drives a push-pull network. In particular, the
ligand-receptor complex catalyzes the phosphorylation of the readout
molecules, while activated readouts can spontaneously decay, see
Fig. 1 in the main text. This system is described by the following
chemical reactions,
\begin{align}
\label{eq:crL}
\mathrm{L+R} & \xrightleftharpoons[k_2]{k_1}  \mathrm{RL} \\
{\rm x} + {\rm RL} & \xrightleftharpoons[k_{-{\rm f}}]{k_{\rm f}}  {\rm x^*} + {\rm RL} \\
{\rm x^*} & \xrightleftharpoons[k_{-{\rm r}}]{k_{\rm r}} {\rm x}
\label{eq:crx}
\end{align}
where L represents the free ligand, R the free receptor, RL the
ligand-bound receptor, ${\rm x}^*$ the activated readout and ${\rm x}$
the deactivated readout. We also assume that the concentrations of
ATP, ADP, and Pi are constant and absorbed in the rate constants. The
cell needs to detect the total concentration $L(t)\equiv{\rm
  [L]_T}(t)$ of ligand molecules, including both free and
receptor-bounded molecules, ${\rm [L]_T}(t)={\rm [L]}(t)+{\rm
  [RL]}(t)$. Moreover, since the total number of receptors $\RT$ is
constant, we can express the number of free receptors as
$R(t)=\RT-RL(t)$. Similarly, the number of unphosphorylated readout
molecules is $x(t)=X_{\rm T}-x^*(t)$, with $X_{\rm T}$ the total
number of readout molecules and $x^*$ the number that is
phosphorylated. Finally, we assume that we can neglect the
sequestration of ligand molecules by the receptors, yielding ${\rm
  [L]}(t) \simeq {\rm [L]_T}(t)=\LT(t)$ (for ease of notation we thus
drop the subscript $_{\rm T}$ on the total ligand concentration
$\LT(t)$).

The chemical Langevin equations for this system read
\begin{align}
\dot{RL}(t) &= k_1 \LT(t)(\RT-RL(t))-k_2RL(t)+\eta_{RL}\\
\dot{x^*}(t) &= k_{\rm f} RL(t) (X_{\rm T}-x^*(t))-k_{-{\rm f}} RL(t)x^*(t)\nonumber \\
&+k_{-{\rm r}}(X_{\rm T}-x^*(t))-k_{\rm r}x^*(t) + \eta_{x^*}
\end{align}
with independent Gaussian white noise functions
\cite{Gillespie2000,Gardiner2009,Warren2006,Tanase-Nicola2006,Walczak2012}.
These equations reduce to the chemical rate equations for large copy
numbers. We then apply the Linear-Noise Approximation (LNA)
\cite{vanKampen1992}: we expand the rate equations to first order
around the steady-state of the mean-field chemical rate equations and
compute the noise strength at this steady state. Comparison with
computer simulations has revealed that when the system fluctuates in
one basin of attraction, this description is surprisingly accurate
even when the average copy numbers are as small as 10 molecules
\cite{Tanase-Nicola2006,Ziv:2007bo}.  In this approximation, the
distribution of copy numbers is given by a multivariate Gaussian
distribution \cite{vanKampen1992}.  It implies that the problem of
computing the {\snr} in Eq. 20 of the main text and thus the mutual
information between the instantaneous values of the input and output
reduces to calculating the variances and covariances of the
corresponding copy numbers \cite{Tostevin2009,Tostevin2010}. We also
emphasize that the external quantity $\LT(t)$ is a concentration,
while the internal quantities $R, RL, x, x^*$ are copy numbers.

We apply the LNA, expanding the ligand concentration and the receptor
and readout copy numbers around their steady-state values as given by
the mean-field chemical rate equations:
$\LT(t)=\overline{\LT}+\delta \LT(t)$, $RL(t)=\overline{RL}+\delta
RL(t)$ and $x^*(t)=\overline{x}^*+\delta x^*(t)$, with mean values
$\overline{RL}= k_1 \overline{\LT} \RT /(k_2+ k_1\overline{L}_{\rm
  T})$ and $\overline{x}^*= (k_{\rm f} \overline{RL}+k_{-{\rm
    r}})X_{\rm T}/((k_{\rm f}+k_{-{\rm f}})\overline{RL}+k_{\rm r}+
k_{-{\rm r}})$. We then consider the Langevin dynamics of the new
variables $\delta \LT(t)$, $\delta RL(t)$ and $\delta x^*(t)$ that
describe the fluctuations around the corresponding mean values,
\begin{align}
  \delta \dot{RL}(t) 
  &= \rho \delta \LT(t)-\mu \delta RL(t)+\eta_{RL} \elabel{dRL}\\
  \delta \dot{x^*}(t) 
  &= \rho^\prime \delta RL(t)-\mu^\prime \delta x^*(t) + \eta_{x^*}.\elabel{ddx}
\end{align}
In the first equation, $\mu=k_1\overline{\LT}+k_2=\tc^{-1}$ is
the inverse of the receptor correlation time $\tc$, $\rho =\RT k_1
(1-p)=p(1-p)\RT\mu /\overline{\LT}$, where $p=\overline{RL}/ \RT =
k_1\overline{\LT} / (k_2+k_1\overline{\LT})=k_1\overline{\LT} /\mu$ is the fraction of ligand-bound
receptors. The second term on the r.h.s. represents the fluctuations
in the ligand receptor binding at constant ligand concentration,
while the first term arises from the fluctuations in the
total ligand concentration. In the second equation, $\rho^\prime=
k_{\rm f} X_{\rm T}(1-f)-k_{-{\rm f}}X_{\rm T}f $ and
$\mu^\prime=(k_{\rm f}+k_{-{\rm f}}) p\RT + k_{\rm r}+k_{-{\rm
    r}}=\tr^{-1} $ is the inverse of the integration time $\tr$, where
$f=\overline{x}^* / X_{\rm T} = (k_{\rm f} p\RT+k_{-{\rm r}}) /
(k_{\rm f}+k_{-{\rm f}}) p\RT + k_{\rm r}+k_{-{\rm r}}) = (k_{\rm f}
p\RT +k_{-{\rm r}})\tr$ is the fraction of phosphorylated readout
molecules. The second term on the r.h.s. of \eref{ddx} represents the
fluctuations in the phosphorylation reaction at constant
number of ligand-bound receptors, while the first term
is due to fluctuations in the number of ligand-bound receptors.

The noise functions are given by \cite{Warren2006}
\begin{align}
\avg{\eta^2_{RL}} &= 2\mu \RT p (1-p) \\
\avg{\eta^2_{x^*}} &= 2\mu^\prime X_{\rm T} f(1-f)
\end{align}
where the cross-correlations
$\avg{\eta_{\LT}\eta_{RL}}=\avg{\eta_{x^*}\eta_{\LT}}=\avg{\eta_{x^*}\eta_{RL}}=0$
are zero because receptor-ligand
binding does not affect the total ligand concentration and the complex
$RL$ acts as a catalyst in the push-pull network
\cite{Tanase-Nicola2006}.

\section{\label{app:SNRx} Estimating the concentration from the number
  of readout molecules $x^*$}

{\bf Dynamic input-output relation} The cell infers the current ligand
concentration $\LT(t)$ from the instantaneous concentration of the
output $x^*(t)$ and by inverting the input-output relation
$\overline{x^*}(\LT)$. Since the ligand concentration fluctuates in
time, and because the system will, in general, not respond instantly
to these fluctuations, the input-output relation that the system must
employ is the {\em dynamic} input-output relation, which yields the
average readout concentration $\overline{x^*} (\LT)$ {\em given that}
the current value of the time-varying signal is $\LT(t)$; here, the
average is not only over the noise sources in the propagation of the
signal from the input $\LT$ to the output $x^*$---the receptor-ligand
binding noise and the readout-phosphorylation noise (see Fig. 2(b)
main text)--- but also over the ensemble of input trajectories that
each have the same current concentration $\LT(t)$ (see Fig. 2(c) main
text) \cite{Tostevin2010,Hilfinger:2011ev,Bowsher2013}. This dynamic
input-output relation differs from the static input-output relation
$\overline{x^*}(\LT_{\rm s})$, which gives the average output
concentration $\overline{x^*}$ for a steady-state ligand concentration
$\LT_{\rm s}$ that does not vary in time (or on a timescale that is
much longer than that of the response). The slope of the dynamic
input-output relation, which is key to the sensing precision, can be
obtained from the Gaussian model discussed below.

{\bf Sensing error} Linearizing $\overline{x^*}(\LT)$ around the mean concentration
  $\overline{\LT}$ and using the rules of error propagation, the
expected error in the concentration estimate is then
\begin{align}
(\delta \hat{\LT})^2 = \frac{\sigma^2_{x^*|\LT}}{\tilde{g}^2_{\LT \to x^*}}.
\elabel{LTerr_SI}
\end{align}
In this expression, $\sigma^2_{x^*|\LT}$ quantifies the width of the
distribution of the output $x^*$ \textit{given} a value of the input
signal $\LT$, while $\tilde{g}_{\LT \to x^*}$ is the dynamic gain,
i.e. the slope of $\overline{x^*}(\LT)$ at $\overline{\LT}$.

{\bf Gaussian statistics} We can obtain the variance
$\sigma^2_{x^*|\LT}$ and the dynamic gain $\tilde{g}_{\LT \to x^*}$
within the Gaussian framework of the linear-noise approximation
\cite{Tostevin2010}.  In the Gaussian model, the distribution of input
values $\LT(t)$ and output values $x^*(t)$ is Gaussian around their
mean values, $\overline{\LT}$ and $\overline{x^*}$,
  respectively. We first define the deviations of $\LT$ and $x^*$ away
  from their mean values, respectively:
\begin{align}
\delta \LT(t) &= \LT(t) - \overline{\LT},\\
\delta x^*(t) &= x^*(t) - \overline{x^*}.
\end{align}
Since the dynamics of both $\LT$ and $x^*$ are stationary processes,
we can choose to omit the explicit dependence on time, and simply
write $\delta \LT(t) = \delta \LT$ and similarly for $x^*$,
Defining the vector ${\bf v}$ with components $\delta
  \LT(t),\delta x^*(t)$, the joint distribution can be written as
\begin{align}
p({\bf v}) = \frac{1}{\sqrt{2\pi^{2N}|{\bf
      Z}|}}\exp\left(-\frac{1}{2}{\bf v}^{\rm T}{\bf Z}^{-1}{\bf
    v}\right)
\elabel{pv}
\end{align}
where ${\bf Z}^{-1}$ is the inverse of the matrix ${\bf Z}$, which has
the following form:
\begin{align}
\elabel{Z}
{\bf Z} = \begin{pmatrix} \sigma^2_{\LT} & \sigma^2_{\LT,x^*}\\
\sigma^2_{\LT,x^*} & \sigma^2_{x^*}\end{pmatrix}.
\end{align}
From \eref{pv} it follows that the conditional distribution of $\delta
x^*$
given $\delta \LT$ is
\begin{align}
p(\delta x^*|\delta \LT) = \frac{1}{(2\pi \sigma^2_{x^*|\LT})^{1/2}}
    \exp\left[-\frac{\left(\delta x^* - \overline{\delta x^*}(\delta
          \LT)\right)^2}{2\sigma^2_{x^*|\LT}}\right].
\elabel{pdx}
\end{align}

{\bf Dynamic gain} In \eref{pdx}, $\overline{\delta x^*}(\delta \LT)$
is the average of the deviation $\delta x^*(\delta \LT) = x^*(\delta
\LT) - \overline{x^*}$ of $x^*$ from its mean $\overline{x^*}$ given
that the input is $\delta \LT=\delta \LT(t)=\LT(t)-\overline{\LT}$; it
describes the dynamic input relation $\overline{x^*}(\LT)$
around $\LT=\overline{\LT}$. It is given by $\overline{\delta x^*}(\delta
\LT) = \sigma^2_{\LT,x^*} / \sigma^2_{\LT} \delta \LT \equiv \dg
\delta \LT$, which defines the dynamic gain:
\begin{align}
\dg = \sigma^2_{\LT,x^*} /\sigma^2_{\LT}\elabel{dgx1}.
\end{align}
Here, $\sigma^2_\LT$ is the variance of the input and
$\sigma^2_{\LT,x^*}$ is the covariance between $\LT$ and $x^*$, which
is derived in Appendix S-A, see \eref{covx}. It shows that the dynamic
gain is
\begin{align}
\tilde{g}_{\LT \to x^*}=\rho \rho^\prime /
((\lambda+\mu)(\lambda+\mu^\prime)) = \dg \rho^\prime \RT /
\mu^\prime.\elabel{dgx2}
\end{align}
In contrast to the macroscopic static gain $g_{\LT \to
  x^*}=d\overline{x^*}/d\LT_{\rm s}$, which characterizes the
transmission of signals $\LT_{\rm s}$ that are constant in time, the
dynamic gain depends both on parameters of the readout system and on
the timescale of the input fluctuations $\tL$. Only in the limit of slowly
time-varying signals ($\tL \gg \tr, \tc$), does the dynamic gain
$\tilde{g}_{\LT \to x^*}$ become equal to the static gain $g_{\LT \to x^*}$

{\bf Conditional variance} In \eref{pdx}, the variance
$\sigma^2_{x^*|\LT}$ is the variance in $x^*$ given that the signal is
$\LT$. It is given by $\sigma^2_{x^*|\LT} = |{\bf Z}|/\sigma^2_{\LT}$ \cite{Tostevin2010},
such that 
\begin{align}
\sigma^2_{x^*|\LT} = \sigma^2_{x^*} - \dg^2 \sigma^2_\LT,
\elabel{sigxLT}
\end{align}
where $\sigma^2_{x^*}$ is the full variance of $x^*$, derived in
Appendix S-A, see \eref{sigmax}. Indeed, in this Gaussian model, the
total variance $\sigma^2_{x^*}$ in the output $x^*$ can be decomposed
into a contribution from the variance $\tilde{g}^2_{\LT \to x^*}
\sigma^2_{\LT}$ due to variations in the signal itself, and a
contribution from the variance $\sigma^2_{x^*|\LT}$ for a given value
of the input $\LT$.  The conditional variance
 $\sigma^2_{x^*|\LT}$ is shaped both by the noise in the propagation
 of the input $\LT$ to the output $x^*$---stochastic receptor-ligand
 binding and noisy readout activation---and by the dynamics of the
 input signal.

{\bf Signal-to-noise ratio (SNR)} The signal-to-noise ratio (SNR) is given by
\begin{align}
{\rm SNR}&=\frac{\sigma^2_{\LT}}{(\delta \hat{L})^2},
\end{align}
as discussed in the main text (see Eq. 3). Combining this expression
with \eref{LTerr_SI} yields the sensing error, the inverse SNR:
\begin{align}
  \elabel{minfoSIa} {\rm SNR}^{-1}=\frac{\sigma^2_{x^*|\LT}}{\tilde{g}^2_{\LT\to x^*}\sigma^2_\LT}=\frac{\sigma^2_{\LT}\sigma^2_{x^*}}{\sigma^4_{\LT,
      x^*}}-1,
\end{align}
where we have used \erefstwo{dgx1}{sigxLT}. Using the expressions for
the variance for $x^*$, \eref{sigmax}, and the covariance between
$\LT$ and $x^*$, \eref{covx}, the {\snr} reads
\begin{align}
  \elabel{minfoSI} {\rm SNR}^{-1}
  &= \frac{(\lambda+\mu)^2(\lambda+\mu')^2}{\rho^2\rho^{\prime^2}\sigma^2_\LT} f(1-f)X_{\rm T} \nonumber \\
  &+ \frac{(\lambda+\mu)^2(\lambda+\mu')^2}{\sigma^2_\LT \mu^\prime(\mu+\mu^\prime)\rho^2} p(1-p)\RT \nonumber \\
  & +\frac{(\lambda+\mu)(\lambda+\mu')(\lambda+\mu+\mu^\prime)}{\mu
    \mu^\prime(\mu+\mu^\prime)} -1.
\end{align}
This expression is difficult to interpret intuitively and impedes an
analysis of the fundamental resources required for sensing. In
contrast, the description of the readout system as a sampling device,
presented in Sec. \ref{app:samplerrors}, yields a much more
illuminating expression for the sensing error, showing how it arises
from a sampling error in estimating the receptor occupancy, set by the
number of samples, their independence and their accuracy, and a
dynamical error, set by the history of the input signal. In
\ref{app:xtoptr} we show explicitly that these expressions are indeed identical.

Lastly, for the Gaussian model employed here, the SNR defined by
\eref{minfoSI}, can be directly related to the mutual information
\cite{Tostevin2010,Shannon1948}:
\begin{align}
I(\LT; x^*)&=-\frac{1}{2}\ln(1-r^2_{\LT,x^*}),\\
&=\frac{1}{2} \ln (1+\text{SNR}),
\elabel{Iinst}
\end{align}
where $r^2_{\LT,x^*}\equiv \sigma^4_{\LT,x^*}/(\sigma^2_{\LT}
\sigma^2_x)$ is the correlation coefficient between input and
output. This measure has also been used to quantify information
transmission via time-varying signals \cite{Brittain:2017hf,Das:2017gh}.

\section{\label{app:samplerrors}Calculating the SNR within the sampling framework}

In this section we derive the main result of our
manuscript, namely the \snr\ given by Eq. 20 of the main text. We
derive this result by viewing the downstream network as a device that
discretely samples the receptor state, first proposed in
Ref. \onlinecite{Govern2014}. The important quantities are the number
of samples, the spacing between them, and the properties of the
signal. The benefit of viewing the network as a sampling device is
that the resulting expression has an intuitive interpretation: the
more samples, the higher the \snr; the further apart they are, the
more independent they are. Moreover, in contrast to the static case,
we see that even when the number of samples is very large, a
systematic error remains when the integration time is finite; this
dynamical error arises naturally within the sampling framework.

We first derive the \snr\ for the irreversible push-pull network in
section \ref{app:SNRirrev_sampl}, and then generalize its expression
to that of the full system in section \ref{app:SNRrev_sampl}.
To help the reader in getting an overview of the derivation, we
introduce several brief {\bf overview paragraphs highlighted in
  bold, which elucidate the structure of the
derivation}.


\subsection{\label{app:SNRirrev_sampl} The SNR for the irreversible
  system derived within the sampling framework}

We present the derivation of the \snr\ within the sampling framework
for the irreversible system, described by the following reactions: 
\begin{align}
\mathrm{L+R} & \xrightleftharpoons[k_2]{k_1}  \mathrm{RL} \\
{\rm RL} + {\rm X} &\overset{k_{\rm f}}
  \rightarrow {\rm RL} + {\rm X}^*\\
{\rm X^*}  &\overset{k_{\rm r}}\rightarrow {\rm X}.
\end{align}
The input signal $L(t)$ is modeled as a stationary signal with mean
$\overline{\LT}$, variance $\sigma^2_{\LT}$, and correlation time
  $\tL=\lambda^{-1}$. The relaxation of the deviation $\delta \LT(t) =
  \LT(t) -
  \overline{\LT}$ from the mean signal $\overline{\LT}$ is thus characterized by
  the correlation function $\avg{\delta \LT(t) \delta
    \LT(t^\prime)}=\sigma^2_{\LT} e^{-\lambda(t-t^\prime)}$.  The
  ligand molecules bind the receptor molecules stochastically with
  receptor correlation time $\tc =k_1 \overline{L}_{\rm
    T}+k_2=\mu^{-1}$. The readout molecules X interact with the
  receptor such that the ligand binding state of the receptor is
  copied into the chemical modification state of the readout. We
  consider the limit that the total number of readout molecules
  $X_{\rm T}$ is large, such that the fraction of phosphorylated
  readout molecules $f=k_{\rm f} p \RT/(k_{\rm f}p\RT+k_{\rm r})$ is
  small and $\overline{x}\simeq X_{\rm T}$. The integration time of
  this system is $\tr=(k_{\rm f}p \RT+k_{\rm r})^{-1}=
  \mu^{\prime^{-1}}$.

We view the downstream readout system as a sampling device that
estimates the average receptor occupancy over the integration time
$\tr$ from the active readout molecules $x^*(\LT(t))=x^*(\LT)$ via
\begin{align}
\hat{p}_{\tr} = \frac{x^*(\LT)}{\mN},
\elabel{ph}
\end{align}
where $\mN$ is the average of the number of samples $N$ taken during
the integration time $\tr$. 

The number of active readout molecule $x^*(t)$ at time $t$ is given by
\begin{align}
x^*(t) = \sum_{i=1}^N n_i (t_i),
\elabel{x*}
\end{align}
where $n_i$ is the state of the $i$th sample, corresponding to the
state of the receptor involved in the $i$th collision at time $t_i<t$:
$n_i(t_i)=1$ if receptor is ligand bound and $n_i(t_i)=0$
otherwise. The total rate at which inactive readout molecules interact
with the receptor---the sampling rate---is given by $r=k_{\rm
  f}\overline{x} \RT \approx k_{\rm f} \xT \RT$ and the average number
of samples obtained during the integration time $\tr$ is
\begin{align}
\mN = k_{\rm f} X_{\rm T} \RT \tr. 
\elabel{mN}
\end{align}
We also note here that the flux of readout molecules is $\dot{n} = r p$
and, using that $f=k_{\rm f} p \RT \tr$, the average number of samples
is also given by $\mN =  f(1-f) \xT / p \approx f \xT / p$.

The cell then estimates the concentration via its estimate of the
receptor occupancy $\ph$ and by inverting the dynamic input-output relation
$p_\tr (\LT)$. Via error propagation this yields the error 
\begin{align}
(\delta \hat{\LT})^2 = \frac{\sigma^2_{\phL}}{\dg^2},
\elabel{ErrConc}
\end{align}
where $\sigma^2_{\phL}$ is the variance in the estimate of $p_\tr$
given the ligand concentration $\LT(t)$, and
$\dg$ is the dynamic gain $\dg \equiv d p_\tr (\LT) / d\LT$. Defining
the \snr\ as ${\rm SNR}=\sigma^2_{\LT} / (\delta \hat{\LT})^2$, where
$\sigma^2_\LT$ is the variance of the ligand concentration, this
yields
\begin{align}
  {\rm SNR^{-1}}= \frac{(\delta \hat{\LT})^2}{\sigma^2_\LT}=\frac{\sigma^2_{\phL}}{\dg^2\sigma^2_{\LT}}.
  \elabel{SNRinverse_sampl}
\end{align}

{\bf Overview} We first derive the dynamic gain and then in the
section {\bf Error in estimating receptor occupancy} the error
$\sigma^2_{\phL}$. 

{\bf Dynamic gain} The dynamic gain quantifies how much a ligand
fluctuation at time $t$, $\delta \LT(t)$, leads to a change $\delta
p_{\tr}$ in the average receptor occupancy $p_\tr=\avg{n(t)}_{\tr}$
over the past integration time $\tr$. The average of the receptor
occupancy is taken by the readout molecules downstream of the
receptor: these molecules at time $t$ provide the samples of the state
of the receptor at the earlier times $t_i$. As shown in
Ref. \onlinecite{Govern2014}, the probability that a readout molecule
at time $t$ provides a sample of the receptor at an earlier time $t_i$
is $p(t_i|{\rm sample})= e^{-(t-t_i)/\tr}/\tr$. Hence, the average
change in the receptor occupancy over the past integration time $\tr$
is
\begin{align}
\delta p_\tr &= E \avg{\delta n(t_i)}_{\delta \LT(t)}\elabel{dntE}\\
&=\int_{-\infty}^tdt_i \avg{\delta
  n(t_i)}_{\delta \LT(t)} \frac{e^{-(t-t_i)/\tr}}{\tr}.
\elabel{ntr}
\end{align}
Here, $E$ denotes the expectation over the sampling times $t_i$,
$\avg{\delta n(t_i)}_{\delta \LT(t)}$ is the average deviation in the
receptor occupancy at time $t_i$, $\avg{\delta n(t_i)}_{\delta
  \LT(t)}\equiv \avg{n(t_i)}_{\delta \LT(t)} - p$, {\em given that the
  ligand concentration at time $t$ is $\delta \LT(t)$}; this average
is taken over receptor-ligand binding noise and the subensemble of
trajectories ending at $\delta \LT(t)$, see Fig. 2(c) of the main
text. We can compute it within the linear-noise approximation:
\begin{align}
\avg{\delta n(t_i)}_{\delta \LT(t)}= \rho_n \int_{-\infty}^{t_i} dt^\prime
\avg{\delta \LT(t^\prime)}_{\delta \LT(t)} e^{-(t_i -
    t^\prime)/\tc},
\elabel{nti}
\end{align}
where $\rho_n = p(1-p) / (\overline{L}_T \tc)$ and $\avg{\delta
  \LT(t^\prime)}_{\delta \LT(t)}$ is the average ligand concentration at
time $t^\prime$ given that the ligand concentration at time $t$ is
$\delta \LT(t)$. It is given by
\cite{Bowsher2013}
\begin{align}
\avg{\delta \LT(t^\prime)}_{\delta \LT(t)}= \delta \LT(t)
e^{-|t-t^\prime|/\tL}.
\elabel{Ltp}
\end{align}
Combining \erefsrange{ntr}{Ltp} yields the following expression
for the average change in the average receptor occupancy $p_\tr$, given
that the ligand at time $t$ is $\delta \LT(t)$:
\begin{align}
\avg{\delta n(t)}_{\delta \LT(t)}^\tr &= \frac{p(1-p)}{\overline{L}_{\rm
    T}}\left(1+\frac{\tc}{\tL}\right)^{-1}\left(1+\frac{\tr}{\tL}\right)^{-1}\delta
\LT(t),\\
&=\dg \delta \LT(t) \elabel{dnt}.
\end{align}
Hence the dynamic gain is
\begin{align}
\dg&= \frac{p(1-p)}{\overline{L}}\left(1+\frac{\tc}{\tL}\right)^{-1}\left(1+\frac{\tr}{\tL}\right)^{-1},
\elabel{dg}\\
&=g_{\LT \to p}\left(1+\frac{\tc}{\tL}\right)^{-1}\left(1+\frac{\tr}{\tL}\right)^{-1}.
\end{align}
The dynamic gain is the average change in the receptor occupancy $p_\tr$ over the past
integration time $\tr$ given that the change in the ligand
concentration at time $t$ is $\delta \LT(t)$. It depends on all the
timescales in the problem, and only reduces to the static gain
$g_{\LT \to p}=p(1-p) / \overline{L}$ when the integration time $\tr$ and
the receptor correlation time $\tc$ are both much shorter than the
ligand correlation time $\tL$. The dynamic gain determines how
much an error in the estimate of $p_\tr$ propagates to  the estimate
of $\LT(t)$.

{\bf Error in estimating receptor occupancy} Using the law of total
variance, the error
$\sigma^2_{\ph|\LT}$ in the estimate of the receptor occupancy $p_\tr$ over
the past integration time $\tr$ is given by
\begin{align}
\elabel{noiseaddv1}
\sigma_{\ph|\LT}^2 =    \text{var} \left[ E  (\phL |N) \right] + E
\left[ \text{var}(\phL | N) \right].
\end{align}
The first term reflects the variance of the mean of $\phL$ given the
number of samples $N$; the second term reflects the mean of the
variance in $\phL$ given the number of samples $N$
\cite{Govern2014}. 

{\bf Overview} We first discuss the first term  and
then the second term on the right-hand side of \eref{noiseaddv1}. The
second term contains two contributions; one combines with the
first term to give rise to the sampling error, while the other yields
the dynamical error of Eq. 20 of the main text.

{\bf Error from stochasticity in number of samples}
The first term of \eref{noiseaddv1} describes the noise that arises
from the stochasticity in the number of samples. It can be
written as
\begin{align}
 \text{var} \left[ E  (\phL |N) \right] = \text{var} \left[
   \frac{1}{\bar{N}}  E  \left( \sum_{i=1}^N n(t_i)
     \middle\vert N \right) \right] \elabel{varE}
\end{align}
where we have dropped the
subscript $i$ on $n_i$ (compare against \eref{x*}) because in
estimating the average receptor occupancy we can focus on a single
receptor. The above average can be written as
\begin{align}
E  \left( \sum_{i=1}^N n(t_i)
     \middle\vert N \right) &= N 
   \overline{E \avg{n(t_i)}}_{\delta \LT(t)},\\
&=N\left ( p +\overline{E \avg{\delta n(t_i)}}_{\delta \LT(t)}\right),\\
&=N\left(p + \overline{\dgs \delta \LT(t)}\right),\\
&=N p. 
\end{align}
Here the angular brackets $\avg{\dots}_{\delta \LT(t)}$ denote an average over the
ligand binding state of the receptor, with the subscript $\delta
\LT(t)$ indicating that the average is to be taken for a given $\delta
\LT(t)$. The expectation $E$ denotes an average over all samples times $t_i$
(see also \eref{dntE}), and the overline indicates an average over $\delta
\LT(t)$. In going from the second to the third line we have used
\eref{dnt}, with $\dgs$ the short-hand notation for $\dgs = \dg$, as
also used below unless stated otherwise. Hence, \eref{varE} becomes
\begin{align}
 \text{var} \left[ E  (\phL |N) \right] &=  
\text{var} \left[ \frac{N}{\mN}  p \right], \\ &=
\frac{p^2}{\mN^2} \text{var} \left[  N  \right],  \\ &=
\frac{p^2}{\mN}\elabel{varMeanN}.
\end{align}
This term is governed by the nature of the sampling process and does
not depend on the statistics of the input signal. It is indeed the same
as that for sensing static concentrations \cite{Govern2014}. 

{\bf Error for fixed number of samples}
The second term of \eref{noiseaddv1} describes the error in the
estimate of $p_{\tr|\LT}$ that arises for a fixed number of
samples. It is given by
\begin{align}
 E\left[ \text{var}(\phL | N) \right] =  E \left[ \frac{N^2}{\mN^2}
  {\rm var} \left(\frac{\sum_{i=1}^N n_i(t_i)}{N} \middle\vert N
  \right)\right]\elabel{varNfix}
\end{align}
In Appendix S-B we show that
\begin{align}
&{\rm var}\left(\frac{\sum_{i=1}^N n_i(t_i)}{N}|N\right)\nonumber\\
&=\frac{p(1-p)}{N} + \overline{E\avg{\delta n_i(t_i)
      \delta n_j(t_j)}}_{\delta \LT(t)} - \dgs^2 \sigma^2_\LT \elabel{vardelta},
\end{align}
where $\delta n_i(t_i) = n_i(t_i) - p$ and $E$ denotes an average over the
sampling times $t_i$.
As we show next, the receptor covariance $\overline{E\avg{\delta
    n_i(t_i) \delta n_j(t_j)}}_{\delta \LT(t)}$ splits into two
contributions, one that together with the first term of
\eref{vardelta} and with \eref{varMeanN} forms the sampling error, and one
that together with the last term of \eref{vardelta}, $-\dgs^2
\sigma^2_\LT$, forms the dynamical error of Eq. 20 of the main text.

{\bf The receptor covariance} To derive the receptor covariance
$\overline{E\avg{\delta n_i(t_i) \delta n_j(t_j)}}_{\delta \LT(t)}$,
the second term of \eref{vardelta},
we note that the deviation $\delta n_i(t_i)= n_i(t_i) - p$ of the
receptor occupancy $n_i(t_i)$ from the mean $p$ is
\begin{align}
\elabel{niti}
\delta n_i(t_i) =  \int_{-\infty}^{t_i}dt^\prime e^{-(t_i-t^\prime)/\tc} [\rho_n\, \delta \LT(t^\prime) + \xi_i(t^\prime)],
\end{align}
where $\xi_i(t^\prime)$ models the ligand-binding noise of the receptor $i$
at time $t^\prime$.  The covariance for a given $\delta \LT(t)$ is
then given by the sum of two contributions,
\begin{align}
&\avg{\delta n_i(t_i)\delta n_j(t_j)}_{\delta \LT(t)} = \nonumber\\
&\underbrace{\rho_n^2\int_{-\infty}^{t_i} dt^\prime \int_{-\infty}^{t_j}
  dt^\dprime e^{-(t_i - t^\prime)/\tc} \avg{\delta \LT(t^\prime) \delta
    \LT(t^\dprime)}_{\delta \LT(t)} e^{-(t_j - t^\dprime)/\tc}}_\text{covS} \nonumber\\
&+ \underbrace{\int_{-\infty}^{t_i} dt^\prime \int_{-\infty}^{t_j} dt^\dprime e^{-(t_i-t^\prime)/\tc} \avg{\xi_i(t^\prime) \xi_j(t^\dprime)} e^{-(t_j - t^\dprime)/\tc}}_\text{covR}.
\elabel{cov}
\end{align}
Hence, the receptor covariance averaged over  $\delta
\LT(t)$ and the sampling times is
\begin{align}
\overline{E\avg{\delta n_i(t_i)\delta n_j(t_j)}}_{\delta \LT(t)} &=
\overline{E[\text{covS}(n_i(t_i),n_j(t_j))]}\nonumber\\
&+\overline{E[\text{covR}(n_i(t_i),n_j(t_j))]}\elabel{MeanCov}.
\end{align}

{\bf Overview} The first term on the right-hand side of \eref{MeanCov}
describes the receptor covariance due to the ligand concentration
fluctuations. Together with the third term of \eref{vardelta},
$-\dgs^2\sigma^2_\LT$, it forms the dynamical error in estimating
$p_\tr$ (Eq. 18 main text). The second term of \eref{MeanCov}
characterizes the correlations in the receptor switching that arise
from the stochastic ligand binding and unbinding. This term forms,
together with \eref{varMeanN} and with the first term of
\eref{vardelta}, the sampling error in estimating $p_\tr$ (Eq. 17
  main text). We will now first show how these three terms yield the
sampling error. We will then return to the first term on the
right-hand side of \eref{MeanCov} and show how that with $-\dgs^2
\sigma^2_\LT$ it forms the dynamical error.

{\bf Receptor switching noise} In \eref{cov}, $\avg{\xi_i(t^\prime)
  \xi_j(t^\dprime)} = \avg{\xi^2} \delta (t^\prime - t^\dprime)$ where
the noise amplitude
$\avg{\xi^2} = 2p(1-p)/(\RT \tc)$ is
divided by $\RT$ because we 
assume that the ligand molecules bind the receptors independently,
thus ignoring spatio-temporal correlations
\cite{tenWolde:2016ih}.  The second term of \eref{cov} then yields
\begin{align}
\text{covR}(n_i(t_i), n_j(t_j)) &= \frac{p(1-p)}{\RT}e^{-|t_j - t_i|/\tc}.
\end{align}
We now perform the averaging over the sampling times, denoted by
$E$. It is convenient to express and evaluate the integrals in terms
of $\lambda = 1 / \tL$, $\mu=1/\tc$, and $\mu^\prime=1/\tr$. Using
that the probability that a readout molecule at time $t$ has taken a
sample of the receptor at an earlier time $t_i$ is $p(t_i|{\rm
  sample}) = e^{-(t-t_i)/\tr}/\tr$ \cite{Govern2014}, we obtain
\begin{align}
&E[{\rm covR}(n_i(t_i), n_j(t_j))] \nonumber\\
&= \frac{p(1-p)\mu^{\prime^2}}{\RT}\ \int_{-\infty}^t dt_i \int_{-\infty}^t dt_j e^{-\mu^\prime(t-t_i)} e^{-\mu^\prime(t-t_j)} e^{-\lambda|t_j - t_i|}\\
 &= \frac{p(1-p)\mu^{\prime^2}}{\RT} e^{-2 \mu^\prime t} \int_{-\infty}^t dt_j
 e^{2 \mu^\prime t_j} 2\int_0^\infty d\tilde{\Delta}
 e^{-(\mu^\prime+\mu)\tilde{\Delta}}\\
&=\frac{p(1-p)}{\RT} \frac{\tc}{\tc+\tr}\simeq \frac{p(1-p)}{\RT} \frac{\tc}{\tr},\elabel{varphrec}
\end{align}
where $\tilde{\Delta}=|t_j - t_i|$ and in the last line we have used
that typically $\tr \gg \tc$. Clearly, the above expression is the same
for each value of the signal $\delta \LT(t)$ and does not need to be
averaged over $\delta \LT(t)$.

{\bf The sampling error}
\eref{varphrec} forms with the first term of
\eref{vardelta} the sampling error for a fixed
number of samples $N$ (see Ref. \onlinecite{Govern2014}):
\begin{align}
{\rm var}\left(\frac{\sum_{i=1}^N n_i(t_i)}{N}\right)^{\rm
  samp}
 &=\frac{p(1-p)}{N}\left(1+\frac{2N\tc}{2\RT\tr}\right),\\
&=\frac{p(1-p)}{f_I N},\elabel{varsampNfix}
\end{align}
where 
\begin{align}
f_I =\frac{1}{1+2\tc / \Delta}
\end{align}
is the fraction of independent samples with $\Delta =2 \RT \tr / N$ being
the spacing between the receptor samples. We now have to average
\eref{varsampNfix} over the different number of samples $N$ (see \eref{varNfix}), which finally gives
\begin{align}
 E\left[ \text{var}(\phL | N) \right]^{\rm samp}=\frac{p(1-p)}{f_I \mN}.\elabel{EvarNfix}
\end{align}
This equation has a very clear interpretation: it is the
error in the estimate of the receptor occupancy based on a single
measurement---given by the variance of the receptor occupancy
$p(1-p)$---divided by the total number of independent measurements
$f_I \mN$.

\eref{varMeanN} and \eref{EvarNfix} together yield the sampling error
in estimating the receptor occupancy
\begin{align}
\sigma^{2,\,\,{\rm samp}}_{\phL}=\frac{p^2}{\mN} + \frac{p(1-p)}{f_I
  \mN}.
\elabel{sampErrp}
\end{align}

To know how the error $\sigma^{2,\,\,{\rm samp}}_{\phL}$ in the
estimate of the receptor occupancy propagates to the error $(\delta
\hat{\LT})^2$ in the estimate of the concentration (see
\eref{ErrConc}), we need to divide this error by the dynamic gain,
given by \eref{dg}. Via \eref{SNRinverse_sampl} this then yields the
inverse \snr\ associated with the sampling error:
\begin{align}
&{\rm SNR}^{-1}_{\rm samp}\nonumber\\
&=\left(1+\frac{\tc}{\tL}\right)^2\left(1+\frac{\tr}{\tL}\right)^2\left[\frac{\left(\overline{\LT}/\sigma_{\LT}\right)^2}{p(1-p)
    f_I \mN} + \frac{\left(\overline{\LT}/\sigma_{\LT}\right)^2}{(1-p)^2 \mN}\right].\elabel{SNRsamp}
\end{align}

{\bf Dynamical error} In estimating a time-varying ligand
concentration, the sensing error arises not only from the stochastic
sampling of the receptor state, but also from the fact that the
current ligand concentration corresponds to an ensemble of ligand
trajectories in the past, which each give rise to a different
integrated receptor occupancy. This effect is contained in the first
term of \eref{MeanCov}. Crucially, the averaging over $\delta \LT(t)$
can be performed before the averaging over the sampling times, such
that $\overline{E[\text{covS}(n_i(t_i),n_j(t_j))]}=E[\overline{\text{covS}(n_i(t_i),n_j(t_j))}]$ with
\begin{align} 
&\overline{\text{covS}(n_i(t_i),n_j(t_j))} =\nonumber \\
&\rho_n^2\int_{-\infty}^{t_i} dt^\prime \int_{-\infty}^{t_j}
  dt^\dprime e^{-(t_i - t^\prime)/\tc} \overline{\avg{\delta \LT(t^\prime) \delta
    \LT(t^\dprime)}}_{\delta \LT(t)} e^{-(t_j - t^\dprime)/\tc}.\elabel{MeanCovS}
\end{align}
We can now exploit that $\overline{\avg{\delta \LT(t^\prime)\delta
    \LT(t^\dprime)}}_{\delta \LT(t)}=\avg{\delta \LT(t^\prime)\delta
    \LT(t^\dprime)}=\sigma^2_\LT e^{-\lambda|t^\dprime -
    t^\prime|}$. Inserting this into \eref{MeanCovS} 
 and integrating it
yields
\begin{align}
\overline{\text{covS}(n_i(t_i), n_j(t_j))}=\rho^2_n  \sigma^2_\LT
\frac{\lambda e^{-\mu(t_j-t_i)}-\mu e^{-\lambda(t_j-t_i)}}{\mu(\lambda^2-\mu^2)}.
\end{align}
We now again average over the sample times
\begin{align}
&E[\overline{\text{covS}(n_i(t_i), n_j(t_j))}]\\
&\mu^{\prime^2}\int_{-\infty}^tdt_i
\int_{-\infty}^t dt_je^{-\mu^\prime (t-t_i)} \overline{\text{covS}(n_i(t_i),
n_j(t_j))} e^{-\mu^\prime(t-t_j)}\\
&=\rho_n^2\sigma^2_\LT
\frac{\mu^\prime(\lambda+\mu+\mu^\prime)}{\mu(\mu+\lambda)(\mu+\mu^\prime)(\mu^\prime+\lambda)}\\
&=\frac{\rho_n^2
\mu^{\prime^2}\sigma^2_\LT}{(\mu+\lambda)^2(\mu^\prime+\lambda)^2}\frac{(\lambda+\mu+\mu^\prime)(\mu+\lambda)(\mu^\prime+\lambda)}{\mu^\prime
\mu(\mu+\mu^\prime)}\\
&=\dgs^2\sigma^2_\LT \frac{(\lambda+\mu+\mu^\prime)(\mu+\lambda)(\mu^\prime+\lambda)}{\mu^\prime
\mu(\mu+\mu^\prime)}\\
&=\dgs^2 \sigma^2_\LT \left(1+\frac{\tc}{\tL}\right)\left(1+\frac{\tr}{\tL}\right)\left(1+\frac{\tc\tr}{\tL(\tc+\tr)}\right).
\end{align}

Importantly, the above expression is not the dynamical error in the
estimate of the receptor occupancy. It is the receptor covariance that
arises from the signal fluctuations, but this contains a contribution
from the dynamical error {\em and} the signal variations of interest,
$\dgs^2 \sigma^2_\LT$. To obtain the dynamical error in the receptor
occupancy, we have to subtract from the above expression $\dgs^2
\sigma^2_\LT$, which is indeed the third term of \eref{vardelta}---the
term that we had not yet taken care-off. This procedures
directly yields the dynamical error in the receptor occupancy, because
the above expression does not depend on the number of samples $N$, so
there is no need to average over $N$ in \eref{varNfix}. We thus
immediately find
\begin{align}
\sigma^{2, \,\, {\rm dyn}}_{\phL}=\dgs^2 \sigma^2_\LT\left[
  \left(1+\frac{\tc}{\tL}\right)\left(1+\frac{\tr}{\tL}\right)\left(1+\frac{\tc\tr}{\tL(\tc+\tr)}\right)-1\right].
\elabel{dynErrp}
\end{align}

To obtain the contribution from the dynamical error to the \snr\ we
divide \eref{dynErrp} by the dynamic gain and the signal variance, see
\eref{SNRinverse_sampl}. This yields
\begin{align}
{\rm SNR}^{-1}_{\rm dyn}= \left(1+\frac{\tc}{\tL}\right)\left(1+\frac{\tr}{\tL}\right)\left(1+\frac{\tc\tr}{\tL(\tc+\tr)}\right)-1.\elabel{SNRdyn}
\end{align}
Interestingly, this contribution only depends on the timescales in the
problem, which can be understood by noting it arises from the
fact that the signal in the past deviates, in general, from the
current signal. It thus neither depends on the number of
receptors nor on the number of readout molecules that store the
receptor state.

{\bf Sensing error} Summing \erefstwo{SNRsamp}{SNRdyn} gives the
sensing error for the irreversible system. In the next subsection, we
show how the principal result of our study, Eq. 20, the sensing error
for the full system, can be cast in precisely the same form.

{\bf Check} We can check the final expression for the sensing error as
derived within the sampling framework, by computing
$\sigma^2_{x^*|\LT}$ in the linear-noise approximation and exploiting
that $\ph = x^*/\mN$ (see \eref{x*}) such that $\sigma^2_{\ph|\LT} =
\sigma^2_{x^*|\LT}/ \mN^2$, with $\mN = k_{\rm f} \xT\RT\tr$. In
section \ref{app:SNRx} we derived $\sigma^2_{x^*|\LT}$ and the
gain $\tilde{g}^2_{\LT \to x^*}$ for the reversible
system. Taking the irreversible limit, $k_{\rm -f}\to 0$ and $k_{\rm
  -r}\to 0$, then yields via ${\rm SNR}^{-1} = \sigma^2_{x^*|\LT} /
(\tilde{g}^2_{\LT \to x^*}\sigma^2_\LT)$ indeed the same result for
the sensing error.


\subsection{\label{app:SNRrev_sampl} The SNR for the general
  reversible system derived within the sampling framework}

We can derive the principal result of the main text,
Eq. 20, for the fully reversible system by exploiting the mapping
\begin{align}
\sigma^2_{\ph|\LT}=\frac{\sigma^2_{x^*|\LT}}{\mN^2}, \elabel{sigmaphx}
\end{align}
where  $\sigma^2_{\ph|\LT}$ is the variance in
the estimate of the receptor occupancy over the integration time
$\tr$, and $\sigma^2_{x^*|\LT}$ is the variance in the number of
phosphorylated readout molecules, both
conditioned on the signal being $\delta \LT(t)$. The conditional
variance (see \eref{sigxLT})
\begin{align}
\sigma^2_{x^*|\LT}=\sigma^2_{x^*} - \tilde{g}^2_{\LT \to x^*}
\sigma^2_{\LT}\elabel{sigmaxLT}
\end{align} 
is the full variance $\sigma^2_{x^*}$ of $x^*$ minus the variance
$\tilde{g}^2_{\LT \to x^*} \sigma^2_{\LT}$ that is due to the signal
variations, given by the dynamic gain $\tilde{g}^2_{\LT \to x^*}$
from $\LT$ to $x^*$ times the signal variance $\sigma^2_\LT$.

The full variance of the readout $\sigma^2_{x^*}$ in \eref{sigmaxLT} is given by
\eref{sigmax} of Appendix S-A, where we derive the variances and
covariances of the ligand, receptor and the readout.  In this expression,
$\mu=\tc^{-1}=k_1 \overline{\LT}+k_2$ is the inverse of the receptor
correlation time $\tc$; $p=\overline{RL}/\RT=k_1 \overline{\LT} / (k_2
+ k_1 \overline{\LT}) = k_1 \overline{\LT} \tc$ is the probability
that a receptor is bound to ligand; $\rho=\RT k_1 (1-p)=p(1-p)\RT\mu
/\overline{\LT}$; $\mu^\prime=(k_{\rm f} +k_{\rm -f}) p \RT + k_{\rm r}
+ k_{\rm -r}=\tr^{-1}$ is the inverse of the integration time $\tr$;
$f=\overline{x^*}/\xT = (k_{\rm f} p \RT + k_{\rm -r})\tr$ is the
fraction of phosphorylated readout; $\rho^\prime=k_{\rm f}
  \xT(1-f)-k_{-\rm f} \xT f = \dot{n} /(p\RT)$ is the sampling rate
  $\dot{n}$ of the receptor, be it ligand bound or not. Moreover, the
quality factor $q=(e^{\Delta \mu_1}-1)(e^{\Delta \mu_2}-1)/(e^{\Delta
  \mu}-1)=\rho^\prime p \RT \tr / (f(1-f) \xT)=\dot{n}\tr/(f(1-f)
\xT)$ (see Appendix S-C).

To get $\sigma^2_{\ph|\LT}$ from \erefstwo{sigmaphx}{sigmaxLT} we need
not only $\sigma^2_x$ (\eref{sigmax}), but also the average number of
samples $\mN$ and the dynamic gain $\tilde{g}^2_{\LT \to x^*}$.  The
average number of samples taken during the integration time $\tr$ is
$\mN= r \tr = \dot{n} \tr / p = f(1-f)\xT q/p=\rho^\prime
\RT/\mu^\prime$ and the effective number of reliable samples is
$\mN_{\rm eff}=q\mN$. Moreover, \eref{dgx2} reveals that the dynamic
gain from $\LT$ to $x^*$ is given by \cite{Tostevin2010}
\begin{align}
\tilde{g}_{\LT \to x^*} = \frac{\sigma^2_{\LT,x^*}}{\sigma^2_\LT}=\dg \RT \frac{\rho^\prime}{\mu^\prime}=\dg \mN.\elabel{dgx}
\end{align}

Hence, combining \erefsrange{sigmaphx}{dgx} with \eref{sigmax} yields
\begin{align}
\sigma^{2}_{\phL} &= \frac{p^2}{\mN_{\text{eff}}}+ \frac{p(1-p)}{\Neff}+ \frac{p(1-p)}{\RT (1+\tr/\tc)}\nonumber\\
&+ \dg^2\sigma^2_\LT \Big(1+\frac{\tc}{\tL} \Big)
\Big(1+\frac{\tr}{\tL} \Big) \Big(1+ \frac{\tc \tr}{\tL(\tc+\tr)}
\Big)\nonumber\\
&-\dg^2 \sigma^2_\LT.
\end{align}
This can be rewritten using the expression for the fraction of
independent samples, which, assuming that $\tr\gg\tc$, is $f_I=1/(1+2
\tc / \Delta)$, with $\Delta = 2 \tr \RT / \mN_{\text {eff}}$ the
effective spacing between the samples:
\begin{align}
&\sigma^{2}_{\phL} = \frac{p^2}{\Neff}+
\frac{p(1-p)}{f_I\Neff}\elabel{sigmaph_sampl}
\\ &+\dgs^2 \sigma^2_\LT\left[
  \left(1+\frac{\tc}{\tL}\right)\left(1+\frac{\tr}{\tL}\right)\left(1+\frac{\tc\tr}{\tL(\tc+\tr)}\right)-1\right].\nonumber
\end{align}
Importantly, this expression has exactly the same form as that for the
irreversible case (Sec.~\ref{app:SNRirrev_sampl}), obtained by
combining \erefstwo{sampErrp}{dynErrp}, but with $\mN$ replaced by
$\mN_{\text {eff}}=q\mN$ and $\overline{N}_{\rm I}=f_I \mN$ by $f_{\rm
  I}\Neff$. This shows that also for the fully reversible case we can view
the readout system as a device that discretely samples the receptor
state to estimate the occupancy, from which the concentration is then inferred.

Combining \eref{sigmaph_sampl} with \erefstwo{SNRinverse_sampl}{dg} finally
yields the principal result of our work, Eq. 20 of the main text.

\section{\label{app:xtoptr} Estimating concentration from $x^*$ is no
  different from estimating it from $p_\tr$: Rewriting Eq. 4 of main
  text into Eq. 20}
Eq. 4 of the main text (\eref{minfoSI} of {\it SI}) can be
rewritten to yield exactly the same expression as Eq. 20 of the main
text, as it must be possible.  To show the equivalence, it is
convenient to exploit that $\rho=p(1-p)\RT\mu/(\overline{\LT})$,
$\mN=\dot{n} \tr/ p=(\rho^\prime/\mu^\prime) \RT$, $q=\rho^\prime p
\RT \tr / (f(1-f) \xT)$ (see Appendix S-C) and to split the first term
on the right-hand side of \eref{minfoSI}:
\begin{align}
&{\rm SNR^{-1}}\nonumber\\
&=\left(1+\frac{\tc}{\tL}\right)^2\left(1+\frac{\tr}{\tL}\right)^2\left[\frac{f(1-f)\xT\mu^{\prime^2}(\overline{\LT}/\sigma_\LT)^2}{\rho^{\prime^2}(p(1-p)\RT)^2}\right.\nonumber\\
&\left.+\frac{(\overline{\LT}/\sigma_\LT)^2}{p(1-p)\RT(1+\tr/\tc)}\right]\nonumber\\
&+
\left(1+\frac{\tc}{\tL}\right)\left(1+\frac{\tr}{\tL}\right)\left(1+\frac{\tc\tr}{\tL(\tc+\tr)}\right)-1,\\
&=\left(1+\frac{\tc}{\tL}\right)^2\left(1+\frac{\tr}{\tL}\right)^2\left[\frac{\left(\overline{\LT}/\sigma_\LT\right)^2}{\mN_{\rm
      eff}(1-p)^2}+\frac{\left(\overline{\LT}/\sigma_\LT\right)^2}{p(1-p) \mN_{\rm eff}}\right.\nonumber\\
&\left.+\frac{\left(\overline{\LT}/\sigma_\LT\right)^2}{p(1-p)\RT(1+\tr/\tc)}\right]\nonumber\\
&+
\left(1+\frac{\tc}{\tL}\right)\left(1+\frac{\tr}{\tL}\right)\left(1+\frac{\tc\tr}{\tL(\tc+\tr)}\right)-1,\elabel{SNRx-1}\\
&=\left(1+\frac{\tc}{\tL}\right)^2\left(1+\frac{\tr}{\tL}\right)^2\left(\frac{\left(\overline{\LT}/\sigma_\LT\right)^2}{\mN_{\rm
      eff}(1-p)^2}+\frac{\left(\overline{\LT}/\sigma_\LT\right)^2}{p(1-p)\mN_{\rm I}}\right)\nonumber\\
&+\left(1+\frac{\tc}{\tL}\right)\left(1+\frac{\tr}{\tL}\right)\left(1+\frac{\tc\tr}{\tL(\tc+\tr)}\right)-1,\elabel{SNRx}
\end{align}
where $\mN_{\rm I}=f_I \mN_{\rm eff}$ is the effective number of
independent samples, with $f_I=1/(1+2\tc / \Delta)$ the fraction of
independent samples (assuming $\tr\gg\tc$) and $\Delta = 2 \tr \RT /
\mN_{\rm eff}$ the spacing between the samples. \eref{SNRx} is indeed
the central result of the main text, Eq. 20.

\section{\label{app:11to7}Rewriting Eq. 20 of the main text as Eq. 23}

Modeling the readout system as a sampling device yields an intuitive
expression for the sensing error, which shows that the error can be decomposed
into a sampling error and a dynamical error (Eq. 20 main text).
However, to identify the fundamental resources limiting the sensing
accuracy, it is helpful to rewrite the \snr\ in terms of collective
variables that illuminate the cell resources. For that, we start from
\eref{SNRx} in the previous section (i.e., Eq. 20 of the
main text) and we take one step backward in its derivation, by
splitting the second term on the right hand side and exploiting the
expression for the effective number of independent samples $\mN_{\rm
  I}=1/(1+2\tc / \Delta) \mN_{\rm eff}$ with $\Delta = 2 \tr \RT
/\mN_{\rm eff}$ (\eref{SNRx-1}).  We then sum up the first two terms
on the right hand side and use that $\mN_{\rm
  eff}=q\mN=q\dot{n}\tr/p$:
\begin{align}
{\rm SNR^{-1}}&\nonumber\\
&=\left(1+\frac{\tc}{\tL}\right)^2\left(1+\frac{\tr}{\tL}\right)^2\left[\frac{\left(\overline{\LT}/\sigma_\LT\right)^2}{\mN_{\rm eff}p(1-p)^2}\right. \nonumber\\
&\left.+\frac{\left(\overline{\LT}/\sigma_\LT\right)^2}{p(1-p)\RT(1+\tr/\tc)}\right]\nonumber\\
&+
\left(1+\frac{\tc}{\tL}\right)\left(1+\frac{\tr}{\tL}\right)\left(1+\frac{\tc\tr}{\tL(\tc+\tr)}\right)-1\\
&=\left(1+\frac{\tc}{\tL}\right)^2\left(1+\frac{\tr}{\tL}\right)^2\left[\frac{(\overline{\LT}/\sigma_\LT)^2}{(1-p)^2 q\dot{n}\tr}\right.\nonumber\\
&\left.+\frac{(\overline{\LT}/\sigma_\LT)^2}{p(1-p)\RT(1+\tr/\tc)}\right]\nonumber\\
&+
\left(1+\frac{\tc}{\tL}\right)\left(1+\frac{\tr}{\tL}\right)\left(1+\frac{\tc\tr}{\tL(\tc+\tr)}\right)-1.\elabel{SNRcollvar}
\end{align}
This is  Eq. 23 of the main text.

\section{\label{app:dynamicgainOpt} The optimal integration time}
To understand the optimal integration time that maximizes the mutual
information, we first write the inverse \snr\
(\eref{SNRx} or Eq. 20 main text) as
\begin{align}
&{\rm SNR}^{-1}=
\underbrace{\frac{1}{\dg^2 \sigma^2_{\LT}}\left[\frac{p}{\mN_{\rm
      eff}}+\frac{p(1-p)}{\RT(1+\tr/\tc)}\right]}_\text{sampling
error} \nonumber\\
&\underbrace{\left(1+\frac{\tc}{\tL}\right)\left(1+\frac{\tr}{\tL}\right)\left(1+\frac{\tc\tr}{\tL(\tc+\tr)}\right)-1}_\text{dynamical
  error},
\elabel{SNR}
\end{align}
where the dynamical gain is given by \eref{dg}, namely $\dg =
p(1-p)/\overline{\LT} / (1+\tc/\tL) / (1+\tr/\tL)$. The first term in
between the square brackets is the error in the estimate of $p_\tr$
that comes from the combined effect of the stochasticity in the number
of samples, $p^2/\mN_{\rm eff}$ (see \eref{varMeanN}), and that which
comes from the instantaneous sampling, $p(1-p)/\mN_{\rm eff}$ (see
\eref{vardelta}), while the second term describes the error coming
from the correlations between the samples (see
\eref{varphrec}). Clearly, decreasing the integration time $\tr$ helps
to reduce the sensing error by increasing the dynamical gain $\dg$:
this reduces the propagation of the error in the estimate of the
receptor occupancy to that in the concentration. Moreover, decreasing
$\tr$ also helps to reduce the dynamical error. On the other hand,
decreasing $\tr$ also compromises the mechanism of time integration by
reducing the number of independent measurements per receptor,
$(1+\tr/\tc)$. The interplay between these three effects gives rise to
an optimal integration time that maximizes the mutual information.

\section{\label{app:OptDes}The Optimal Design}
\fref{optAllocRTXT} shows the mutual information $I(\LT;x^*)$ as a function of the
number of receptors $\RT$ and the number of readout molecules $\xT$
for the optimal design of the system, obeying the resource allocation
principle of Eq. 31 of the main text. The mutual information is
computed via Eq. 23 of the main text, where we have used that
$\dot{n}\tr = q f(1-f) \xT = q \xT / 4$ because $f\to 1/2$ in the
optimal system. Here, $\tr$ and $q$ are both specified
via the optimal allocation principle, Eq. 31 of the main text:
for a given $\xT$ and $\RT$ (and $\tc$ which is kept fixed), the
optimal integration time $\tr^\ast$ is specified via $\xT =
\RT(1+\tr^\ast/\tc)$ while $\xT=\beta \dot{w}\tr^\ast$ (Eq. 31) specifies $q$ via
$q(\Delta \mu)=(e^{\beta \Delta \mu/2}-1)^2/(e^{\beta \Delta \mu}-1)=4
k_{\rm B}T/\Delta \mu$; solving this for $\Delta \mu$,
yields the optimal $\Delta \mu$, $\Delta \mu^{\rm opt}$, and optimal
$q$, $q^{\rm opt}\equiv q(\Delta \mu^{\rm opt})$.  With these
constraints we can rewrite Eq. 23 of the main text as:
\begin{widetext}
\begin{align}
\elabel{minfo_final1_SI}
\text{SNR}^{-1} &= \left(1+\frac{\tc}{\tL}\right)^2
\left(1+\frac{\tr^\ast}{\tL}\right)^2
\left[\underbrace{\frac{\left(\overline{\LT}/\sigma_\LT\right)^2}{p(1-p)\RT(1+\tr^\ast/\tc)}}_\text{receptor
    input noise} +
  \underbrace{\frac{\left(\overline{\LT}/\sigma_\LT\right)^2}{(1-p)^2q^{\rm
        opt, 2}
      \xT/4}}_\text{coding noise}\right]\nonumber \\
&  +\underbrace{\left(1+\frac{\tc}{\tL}\right) \left(1+\frac{\tr^\ast}{\tL}\right) \left(1+\frac{\tc \tr^\ast}{\tL(\tc+\tr^\ast)}\right) -1}_\text{dynamical error},
\end{align}
\end{widetext}
where $\tr^\ast$ is thus specified via $\xT = \RT(1+\tr^\ast/\tc)$.

\begin{figure}
  \centering
  \includegraphics[width=\columnwidth]{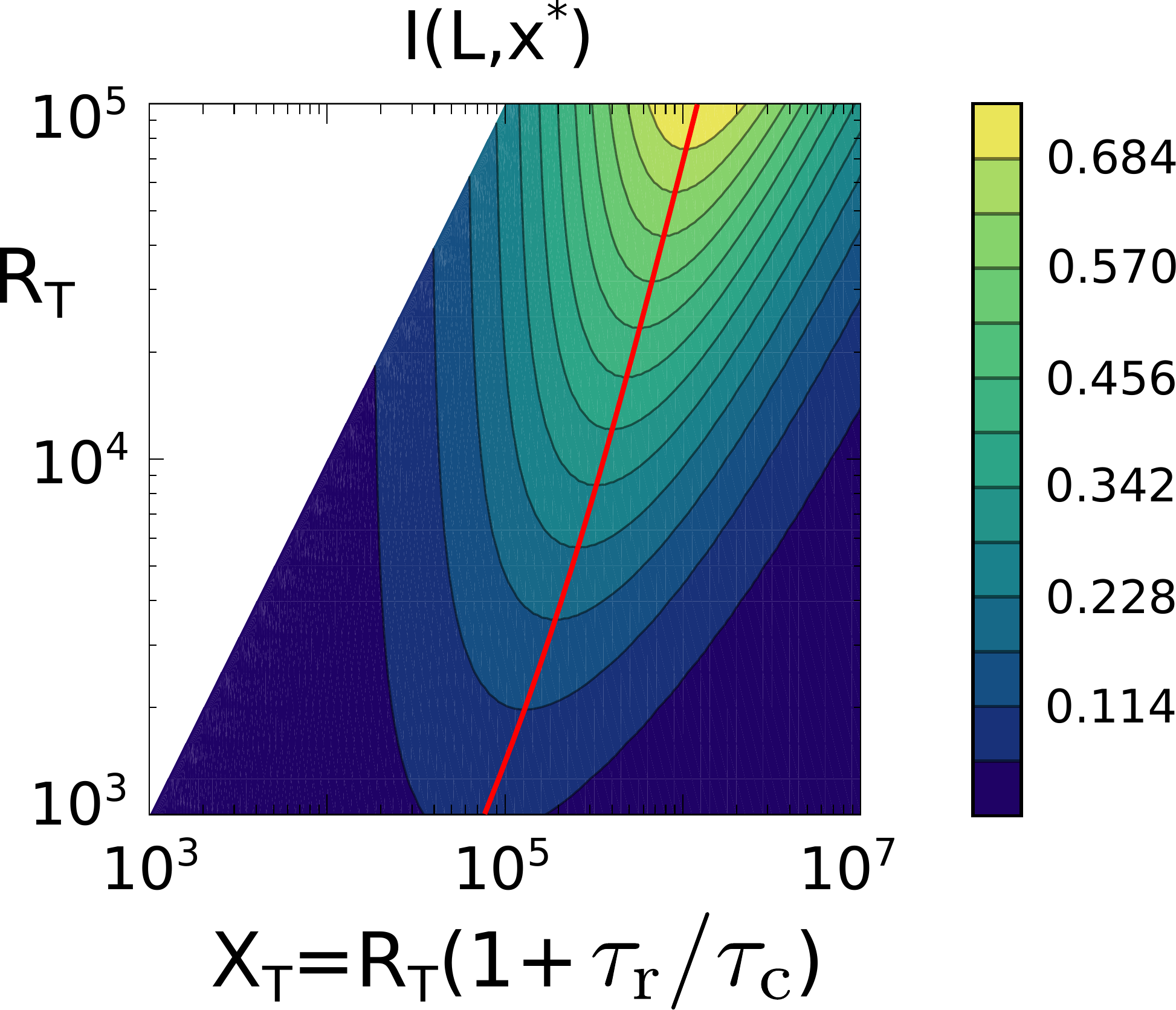}
  \caption{\flabel{optAllocRTXT} The mutual information
    $I(\LT;x^\ast)$ computed via \eref{minfo_final1_SI} as a function
    of $\RT$ and $\xT$ for the optimal system that obeys the optimal
    resource allocation principle Eq. 31 of the main text:
    $\RT(1+\tr^\ast/\tc) = \xT = \beta \dot{w}\tr^\ast$. It is seen
    that for a given $\xT$, the mutual information is maximized for
    $\RT=\xT$. This corresponds to $\tr^\ast=0$, which minimizes the
    dynamical error: the system has given up on time
    integration. However, in this limit, the power diverges. In fact,
    an equilibrium sensing system would be superior, because it can
    reach, for the same number of receptors, the same sensing
    precision as a non-equilibrium one that does not time integrate,
    without turning over costly fuel \cite{Govern2014b}. The
    observation that the membrane is very crowded suggests that the
    number of receptors is limiting. This leads to the design
    principle of Eq. 32 of the main text: $\RT(1+\tr^{\rm
      opt}/\tc) = \xT^{\rm opt} = \beta \dot{w}^{\rm opt}\tr^{\rm
      opt}$; here, the optimal integration time $\tr^{\rm opt}$ is
    computed for a given $\RT$ assuming $\RT$ is limiting (Eq. 24
      main text), while the optimal number of readout molecules
    $\xT^{\rm opt}$ and the optimal power $\dot{w}^{\rm opt}$ are then
    set by this equation. This optimizes the design of the network for
    a given number of receptors $\RT$: $\xT^{\rm opt}$ and
    $\dot{w}^{\rm opt}$ are adjusted to $\RT$ such that they are
    neither in excess nor limiting. Indeed, the contour plot above
    shows that for a given $\RT$ there is an optimal $\xT$, indicated
    by the red line; this value of $\xT$ is close to that given by
    $\xT^{\rm opt}=\RT(1+\tr^{\rm opt}/\tc)$ because $\tr^{\rm opt}$,
    which is computed assuming $\xT$ is in excess ($\xT\to \infty$),
    is close to the optimal $\tr$ that is computed assuming $\xT
    \approx \RT(1+\tr/\tc)$ (the resources limit sensing like weak
    links in a chain, see also Fig. 3(a)). Parameters:
    $\tc/\tL=10^{-2}$; $\sigma_L/\overline{L}_T = 10^{-2}$; $p$
    optimized; $q^{\rm opt}$ given by $q(\Delta \mu) = 4 k_{\rm
      B}T/\Delta \mu$ such that $\xT=\beta \dot{w}\tr^\ast$.}
\end{figure}

\fref{optAllocRTXT} shows the mutual information $I(\LT;x^\ast)$
computed via \eref{minfo_final1_SI} as a function of $\RT$ and $\xT$,
with $\tr^\ast$ specified via $\xT = \RT(1+\tr^\ast/\tc)$. It is seen
that for a given $\xT$ the mutual information is maximized for
$\RT=\xT$. This corresponds to $\tr^\ast=0$: the system has become an
instantaneous responder and has given up on time integration. In this
limit the power diverges. In fact, an equilibrium sensing precision
would be superior, because it can reach the same precision but does
not need to burn fuel \cite{Govern2014b}. The observation that the
membrane is highly packed suggests that sensing systems employ time
integration because $\RT$ is limiting. This leads to the design
principle of Eq. 32 of the main text, $\RT(1+\tr^{\rm
  opt}/\tc)=\xT^{\rm opt}=\beta \dot{w}^{\rm opt}\tr^{\rm opt}$, where
$\tr^{\rm opt}$ is computed in the $\RT$-limiting regime (i.e. via
Eq. 24 main text). This design principle maximizes the sensing
precision for a given $\RT$, and minimizes the number of readout
molecules $\xT$ and the power $\dot{w}$ needed to reach that
precision.

\section*{Appendix S-A: The variances and co-variances}
In the Fourier space we obtain as solutions of \erefstwo{dRL}{ddx}
\begin{align}
\elabel{deltaRL}
\delta \tilde{RL}(\omega) &= {\tilde{\eta}_{RL}\over \mu+i\omega} + {\rho \delta \tilde{\LT}(\omega) \over \mu+i\omega}\\
\delta \tilde{x^*}(\omega) &= {\tilde{\eta}_{x^*}\over \mu^\prime+i\omega} + {\rho^\prime \delta \tilde{RL}(\omega) \over \mu^\prime + i\omega}.
\end{align}
The corresponding power spectra $S_y (\omega)=(1/2T)
\avg{\delta\tilde{y}(\omega)\delta\tilde{y}(-\omega)}$ of $y=\LT,RL,x^*$ are then
given by the spectral addition rule
\cite{Warren2006,Tanase-Nicola2006}
\begin{align}
S_{\LT}(\omega) 
&= {2\lambda \sigma^2_{\LT} \over \lambda^2+\omega^2}\\
S_{RL}(\omega) 
&= {\avg{\eta^2_{RL}} \over \mu^2+\omega^2} +{\rho^2 S_{\delta \LT}(\omega) \over \mu^2+\omega^2}\\
S_{x^*}(\omega) 
&={\avg{\eta^2_{x^*}} \over \mu^{\prime^2} + \omega^2} + {\rho^{\prime^2} S_{\delta RL}(\omega) \over \mu^{\prime^2} + \omega^2},
\end{align}
while the cross power spectra $S_{yz}(\omega)=1/(2T)\avg{\delta \tilde{y}(\omega)\delta\tilde{z}(-\omega)}$ are
\begin{align}
  S_{\LT,RL}(\omega) 
  &= \frac{2k\rho}{(\lambda^2+\omega^2)(\mu-i\omega)}\\
  S_{\LT,x^*}(\omega) 
  &=\frac{2k\rho \rho^\prime}{(\lambda^2 +\omega^2)(\mu-i\omega)(\mu^\prime-i\omega)}\\
  S_{RL,x^*}(\omega) 
&= \frac{\rho^\prime \avg{\eta^2_{RL}}}{(\mu^2+\omega^2)(\mu^\prime-i\omega)}\nonumber \\
&+ \frac{2k\rho^2\rho^\prime}{(\mu^2+\omega^2)(\lambda^2 +\omega^2)(\mu^\prime-i\omega)}.
\end{align}

We can now compute the noise strengths by integration,
$\sigma^2_{yz}=1/(2\pi)\int_{-\infty}^{\infty}d\omega
S_{yz}(\omega)$, resulting in the correlation matrix
\[ \Sigma= \left( \begin{array}{ccc}
    \sigma^2_{\LT} & \sigma^2_{\LT, RL} & \sigma^2_{\LT, x^*} \\
    \sigma^2_{\LT, RL} & \sigma^2_{RL} & \sigma^2_{RL, x^*} \\
    \sigma^2_{\LT, x^*} &\sigma^2_{RL, x^*} &
    \sigma^2_{x^*} \end{array} \right)\] with matrix elements
\begin{align}
\elabel{sigmax}
\sigma^2_{x^*} &= f(1-f)X_{\rm T}+\frac{\rho^{\prime^2}}{\mu^\prime(\mu+\mu^\prime)} \Big[p(1-p)\RT \nonumber \\
& +\frac{\rho^2\sigma^2_\LT (\lambda+\mu+\mu^\prime)}{\mu(\lambda+\mu)(\lambda+\mu^\prime)} \Big],\\
\elabel{sigmaRL}
\sigma^2_{RL} &= p(1-p)\RT+{\rho^2 \sigma^2_\LT \over \mu (\mu+\lambda)}, \\
\elabel{ecovRL}
\sigma^2_{\LT, RL} &= {\rho \sigma^2_\LT \over (\mu+\lambda)}, \\
\elabel{covx} \sigma^2_{\LT, x^*} &= {\rho \rho^\prime \sigma^2_\LT
  \over
  (\lambda+\mu)(\lambda+\mu^\prime)}, \\
\sigma^2_{RL, x^*} &= \frac{\rho^\prime}{\mu^\prime+\mu}\Big[p(1-p)\RT
+{\rho^2\sigma^2_\LT(\mu^\prime+\lambda+\mu) \over \mu
  (\lambda+\mu)(\lambda+\mu^\prime)}\Big].
\end{align}

\section*{Appendix S-B: Variance in average receptor occupancy for fixed
  number of samples}
\eref{vardelta} gives the variance in the average receptor occupancy
for a fixed number of samples. To derive it, we note that

\begin{align}
&{\rm var}\left(\frac{\sum_{i=1}^N n_i(t_i)}{N} \middle\vert N
  \right)_{\delta \LT(t)}\\
&= \frac{\overline{E\avg{(\sum_{i=1}^N n_i
      (t_i))^2}}_{\delta \LT(t)}-
\overline{E\avg{\sum_{i=1}^N n_i (t_i)}^2}_{\delta \LT(t)}}{N^2}\\
&= \frac{\overline{E\avg{(\sum_{i=1}^N n_i
      (t_i))^2}_{\delta \LT(t)}} - N^2 \overline{(p+\dgs \delta
    \LT(t))^2}}{N^2}\elabel{var1}\\
&= \frac{N\overline{(p+\dgs \delta \LT(t))}+N(N-1)\overline{E\avg{ n_i
      (t_i) n_j(t_j)}}_{\delta \LT(t)}}{N^2}\nonumber \\
& - \overline{(p+\dgs \delta
    \LT(t))^2}\elabel{var2}\\
&=\frac{N\overline{(p+\dgs \delta \LT(t))}-N\overline{(p+\dgs \delta
    \LT(t))^2}}{N^2}\nonumber\\
&+\frac{N(N-1)\overline{E\avg{\tilde{\delta} n_i(t_i)
      \tilde{\delta} n_j(t_j)}}_{\delta \LT(t)}}{N^2}\\
&=\frac{p(1-p)- \dgs^2 \sigma^2_\LT}{N} + \frac{N(N-1)\overline{E\avg{\tilde{\delta} n_i(t_i)
      \tilde{\delta} n_j(t_j)}}_{\delta \LT(t)}}{N^2}\elabel{vartil}
\end{align}
Here $\tilde{\delta} n_i(t_i) \equiv n_i(t_i) -
(p+\avg{n_i(t_i)}_{\delta \LT(t)}$ is the deviation away from the
average receptor occupancy $p+\avg{n_i(t_i)}_{\delta \LT(t)}$ when the
ligand concentration at time $t$ is $\delta \LT(t)$. Here, 
$E$ denotes an average over the sampling times of the receptor. The
average of $n_i(t_i)$ over the sampling times $t_i$ given $\delta
\LT(t)$, is $E[\avg{n_i(t_i)}_{\delta \LT(t)}] = p+\dgs \delta \LT(t)$
(see \erefstwo{nti}{dnt}). In addition, $\sigma^2_\LT = \avg{\delta
  \LT(t)^2}$ is the variance of the ligand concentration, and in going
from \eref{var1} to \eref{var2} we have exploited that $n^2=n$. The
quantity $\overline{E\avg{\tilde{\delta} n_i(t_i) \tilde{\delta}
    n_j(t_j)}}_{\delta \LT(t)}$ is the co-variance of the receptor
occupancy given that the ligand concentration at time $t$ is $\delta
\LT(t)$, and then averaged over all values of $\delta \LT(t)$, as
indicated by the overline. This quantity has a contribution from the
receptor switching noise and the dynamical error resulting from the
ligand fluctuations.

While $\overline{E\avg{\tilde{\delta} n_i(t_i) \tilde{\delta}
    n_j(t_j)}}_{\delta \LT(t)}=\overline{E\avg{n_i(t)
    n_j(t_j)}_{\delta \LT(t)}}-\overline{E\avg{n_i}_{\delta
    \LT(t)}\avg{n_j}_{\delta \LT(t)}}$ is the quantity that we need to
compute, it is difficult to compute straightforwardly. We can however
exploit the following trick \cite{Bowsher2013}. Denoting $x=n_i(t_i)$,
$y=n_j(t_j)$ and $z=\delta \LT(t)$, we can write the quantity of
interest as $\overline{E\avg{\tilde{\delta} n_i(t_i) \tilde{\delta}
    n_j(t_j)}}_{\delta \LT(t)}=\overline{E\avg{x
    y}_z}-\overline{E\avg{x}_zE\avg{y}_z}$. We can then
exploit the following
relation
\begin{align}
\overline{E\avg{\delta x \delta y}}_z &=\overline{E\avg{x y}}_z -
  \overline{E\avg{x}_zE\avg{y}_z} \nonumber\\
&+
\overline{E\avg{x}_zE\avg{y}_z} - \overline{E\avg{x}_z}\,\,\overline{E\avg{y}_z},\elabel{dynsigdec}\\
&=\overline{E\avg{x y}}_z -
  \overline{E\avg{x}_zE\avg{y}_z} + \dgs^2 \sigma^2_\LT
\end{align}
Importantly, $\overline{E\avg{\delta x \delta y}}_z$ is the
co-variance related to the deviation of the receptor occupancy from
the mean $p$, which is easier to compute that the deviation from
$p+\dgs\delta \LT(t)$. Inserting the above result into \eref{vartil}
yields
\begin{align}
&{\rm var}\left(\frac{\sum_{i=1}^N n_i(t_i)}{N}\right)_{\delta \LT(t)}\nonumber\\
&=\frac{p(1-p)}{N} + \overline{E\avg{\delta n_i(t_i)
      \delta n_j(t_j)}}_{\delta \LT(t)} - \dgs^2 \sigma^2_\LT \elabel{vardeltaApp}
\end{align}
where we have used that $N(N-1) \approx N^2$ for $N\gg 1$.

\section*{Appendix S-C: The reliability factor $q$}
Here we derive the expression for the reliability factor $q$ in terms
of the fundamental rate constants, $k_{\rm f}$, $k_{\rm -f}$, $k_{\rm
  r}$, $k_{\rm -r}$, which is used to derive the principal result of
the main text, Eq. 20 (i.e. \eref{SNRx} above). We first note that
\begin{align}
  e^{\beta \Delta \mu_1}-1 &= \frac{k_{\rm f} \overline{x}}{k_{\rm
      -f}\overline{x^*}}-1\nonumber\\
&=\frac{k_{\rm f}(k_{\rm r}+k_{\rm -f}pR_T)}{k_{\rm -f}(k_{-\rm
    r}+k_{\rm f}pR_T)}-1 = \frac{k_{\rm f}k_{\rm r}-k_{\rm -f}k_{\rm
    -r}}{k_{\rm -f}(k_{\rm -r}+k_{\rm f}pR_T)}.\\
    e^{\beta \Delta \mu_2}-1 &= \frac{k_{\rm r} \overline{x^*}}{k_{\rm
      -r}\overline{x}}-1\nonumber\\
 &= \frac{k_{\rm r}(k_{\rm -r}+k_{\rm
        f}pR_T)}{k_{\rm -r}(k_{\rm r}+k_{\rm -f}pR_T)}-1 =
    \frac{(k_{\rm f}k_{\rm r}-k_{\rm -f}k_{\rm -r})pR_T}{k_{\rm
        -r}(k_{\rm r}+k_{\rm -f}pR_T)}.
\end{align}
Noting that $\Delta \mu = \Delta \mu_1 + \Delta \mu_2$, we find that
\begin{align}
    e^{\beta \Delta \mu}-1 &= \frac{k_{\rm f}k_{\rm r}}{k_{\rm
        -f}k_{\rm -r}}-1 = \frac{k_{\rm f}k_{\rm r}-k_{\rm -f}k_{\rm
        -r}}{k_{\rm -f}k_{\rm -r}}.
\end{align}
Combining the above results yields the reliability factor $q$:
\begin{align}
    q &\equiv \frac{(e^{\beta \Delta \mu_1}-1)(e^{\beta \Delta
        \mu_2}-1)}{e^{\beta \Delta \mu}-1}\nonumber\\
&=\frac{(k_{\rm f}k_{\rm
        r}-k_{\rm -f}k_{\rm -r})pR_T}{(k_{\rm -r}+k_{\rm
        f}pR_T)(k_{\rm r}+k_{\rm -f}pR_T)}\elabel{qk}
\end{align}
In order to obtain an expression for the reliability factor $q$ that does
not explicitly depend on the rate constants, we notice that the flux of
readout activation $\dot{n}$ can be expressed as
\begin{align}
  \dot{n} &\equiv k_{\rm f}\overline{RL}\overline{x}-k_{\rm -f}\overline{RL}
  \overline{x^*} \nonumber\\
  &= k_{\rm f}p\RT(1-f)\xT-k_{\rm -f}p\RT f\xT\nonumber\\
  &= p\RT\xT[k_{\rm f}(1-f)-k_{\rm -f}f]\nonumber\\
  &= p\RT\xT[k_{\rm f}(k_{\rm -f}p\RT+k_{\rm r})\tr
    -k_{\rm -f}(k_{\rm f}p\RT+k_{\rm -r})\tr]\nonumber\\
  &= p\RT\xT\tr(k_{\rm f}k_{\rm r}-k_{\rm -f}k_{\rm -r}),
\end{align}
where in the forth line we have exploited that $f = (k_{\rm f}p\RT+k_{\rm -r})\tr$.
Substituting $(k_{\rm f}k_{\rm r}-k_{\rm -f}k_{\rm -r})p
\RT=\dot{n}/(\xT \tr)$ into \eref{qk}
and exploiting again the expression for $f$ yields for the
reliability factor $q$
\begin{align}
    q &=\frac{\dot{n}\tr}{f(1-f)X_T}.
\end{align}
This expression shows that the flux $\dot{n}$ can also be expressed as
a function of $q$, namely
\begin{align}
  \dot{n} &= f(1-f)\xT q/\tr.
\end{align}
Finally, the reliability factor $q$ can also be expressed in terms
of the variable $\rho'=k_{\rm f}(1-f)\xT-k_{\rm -f}f\xT$ by
rewriting the flux $\dot{n}$ as
\begin{align}
  \dot{n} &= p\RT\xT[k_{\rm f}(1-f)-k_{\rm -f}f]=\rho' p \RT,
\end{align}
yielding for $q$
\begin{align}
    q &=\frac{\rho' p \RT\tr}{f(1-f)X_T}.
\end{align}

\bibliographystyle{unsrt}

\end{document}